\documentclass[12pt]{article}

\usepackage[utf8x]{inputenc}      
\usepackage{hyperref}
\usepackage{amsmath,amssymb,mathtools}
\usepackage[T1]{fontenc}          
\usepackage{booktabs,tabularx}
\usepackage{graphicx,subfig}
\usepackage{xspace}
\usepackage[usenames]{xcolor}\definecolor{fscolor}{RGB}{44,118,255}
\usepackage{geometry}
\usepackage{todonotes}
\usepackage{listings}
\usepackage[absolute]{textpos}
\usepackage{xparse}
\usepackage[font=small,labelfont=bf,format=plain,margin=0.05\textwidth]{caption}
\usepackage{bbm}
\usepackage{tabularx}
\usepackage{cite}
\usepackage{soul}

\allowdisplaybreaks

\evensidemargin \oddsidemargin
\newcommand{\DR}{{\ensuremath{\overline{\text{DR}}}}\xspace}
\newcommand{\MS}{{\ensuremath{\overline{\text{MS}}}}\xspace}
\newcommand{\OS}{{\text{OS}}\xspace}

\newcommand{\SUSY}{{\text{SUSY}}}

\newcommand{\FH}{\mbox{{\tt FeynHiggs}}\xspace}

\newcommand{\Fig}[1]{Fig.~\ref{#1}}
\newcommand{\Sec}[1]{Section~\ref{#1}}

\newcommand{\Eq}[1]{Eq.~(\ref{#1})}

\newcommand{\cp}{\ensuremath{{\cal CP}}}

\newcommand{\msusy}{\ensuremath{M_\SUSY}\xspace}

\newcommand{\xtDR}{\ensuremath{X_t^\DR/\msusy}\xspace}
\newcommand{\xtOS}{\ensuremath{X_t^\OS/\msusy}\xspace}

\newcommand{\tev}{\,\, \mathrm{TeV}}
\newcommand{\gev}{\,\, \mathrm{GeV}}
\newcommand{\mev}{\,\, \mathrm{MeV}}

\newcommand{\order}[1]{\ensuremath{{\cal O}(#1)}}
\newcommand{\al}{\alpha}
\newcommand{\als}{\al_s}
\newcommand{\alt}{\al_t}
\newcommand{\alb}{\al_b}

\newcommand{\Hpm}{{H^\pm}}

\begin{document}

\thispagestyle{empty}
\def\thefootnote{\fnsymbol{footnote}}

\begin{flushright}
DESY-19-083\\
IFT-UAM/CSIC-19-076\\
MPP-2019-242
\end{flushright}
\vspace{3em}
\begin{center}
{\Large\bf Theoretical uncertainties in the MSSM Higgs boson\\[.5em]
mass calculation}
\\
\vspace{3em}
{
Henning Bahl$^a$\footnote{email: henning.bahl@desy.de},
Sven Heinemeyer$^{b,c,d}$\footnote{email: sven.heinemeyer@cern.ch},
Wolfgang Hollik$^e$\footnote{email: hollik@mpp.mpg.de},
Georg Weiglein$^a$\footnote{email: georg.weiglein@desy.de}
}\\[2em]
{\sl ${}^a$Deutsches Elektronen-Synchrotron DESY, Notkestra{\ss}e 85, D-22607 Hamburg, Germany}\\
{\sl ${}^b$Instituto de F\'isica Te\'orica, (UAM/CSIC), Universidad Aut\'onoma de Madrid, Cantoblanco, E-28049 Madrid, Spain}\\
{\sl ${}^c$Campus of International Excellence UAM+CSIC, Cantoblanco, E-28049 Madrid, Spain}\\
{\sl ${}^d$Instituto de F\'isica  Cantabria (CSIC-UC), E-39005 Santander, Spain}\\
{\sl ${}^e$Max-Planck Institut f\"ur Physik, F\"ohringer Ring 6, D-80805 M\"unchen, Germany}\\
\def\thefootnote{\arabic{footnote}}
\setcounter{page}{0}
\setcounter{footnote}{0}
\end{center}
\vspace{2ex}
\begin{abstract}
{}

The remaining theoretical uncertainties from unknown higher-order corrections in the prediction for the light Higgs-boson mass of the MSSM are estimated. The uncertainties associated with three different approaches that are implemented in the publicly available code \FH are compared: the fixed-order diagrammatic approach, suitable for low SUSY scales, the effective field theory (EFT) approach, suitable for high SUSY scales, and the hybrid approach which combines the fixed-order and the EFT approaches. It is demonstrated for a simple single-scale scenario that the result based on the hybrid approach yields a precise prediction for low, intermediate and high SUSY scales with a theoretical uncertainty of up to $\sim 1.5\gev$ for large stop mixing and $\sim 0.5\gev$ for small stop mixing. The uncertainty estimate of the hybrid calculation approaches the uncertainty estimate of the fixed-order result for low SUSY scales and the uncertainty estimate of the EFT approach for high SUSY scales, while for intermediate scales it is reduced compared to both of the individual results. The estimate of the theoretical uncertainty is also investigated in scenarios with more than one mass scale. A significantly enhanced uncertainty is found in scenarios where the gluino is substantially heavier than the scalar top quarks. The uncertainty estimate presented in this paper will be part of the public code \FH.

\end{abstract}

\newpage
\def\thefootnote{\arabic{footnote}}


\section{Introduction}
\label{sec:01_intro}

The discovery of a Higgs boson at the LHC~\cite{Aad:2012tfa,Chatrchyan:2012xdj} and the subsequent precise determination of its properties~\cite{Khachatryan:2016vau,ATLAS:TWiki,CMS:TWiki} provide important constraints on models beyond the Standard Model (SM). In many beyond SM (BSM) theories the Higgs-boson mass is a free parameter like in the SM. Consequently, only the Higgs-boson couplings, which were measured to be SM-like, can be used to constrain the BSM parameter space. In supersymmetric (SUSY) models, however, the SM-like Higgs-boson mass itself can be calculated in terms of the model parameters allowing for a very sensitive constraint on the SUSY parameter space.

The most common SUSY model is the Minimal Superymmetric Standard Model (MSSM) \cite{Nilles:1983ge,Haber:1984rc}, which associates a superpartner with each SM degree of freedom. Moreover, it adds a second Higgs doublet with respect to the SM resulting in five physical Higgs bosons. At the tree-level, these are the \cp-even $h$ and $H$ bosons, the \cp-odd $A$ boson and the charged $\Hpm$ bosons. We assume the $h$ boson to be the SM-like Higgs boson discovered at the LHC in this work.\footnote{The possibility of the $H$ boson being SM-like is explored for instance in~\cite{Heinemeyer:2011aa,Bechtle:2012jw,Bechtle:2016kui,Haber:2017erd,Bahl:2018zmf}.} In order to take full advantage of the experimental precision reached in the mass measurement of the Higgs boson, the uncertainty of the theoretical prediction for $M_h$ should be of the same order or even smaller.

Consequently, much work has been invested to evaluate the relevant quantum corrections. In the most direct diagrammatic fixed-order approach, full one-loop, the dominant two-loop as well as partial three-loop corrections have been calculated (for recent works see~\cite{Borowka:2015ura,Goodsell:2016udb,Passehr:2017ufr,Harlander:2017kuc,Borowka:2018anu,R.:2019ply,Goodsell:2019zfs}). This fixed-order approach, however, contains large logarithmic contributions, which can limit the accuracy of the perturbative expansion in case of a high SUSY scale. Effective field theory (EFT) techniques have been used to resum the large logarithms providing a precise prediction for large SUSY scales (for recent works see~\cite{Hahn:2013ria,Draper:2013oza,Bagnaschi:2014rsa,Vega:2015fna,Lee:2015uza,Bagnaschi:2017xid,Bahl:2018jom,Harlander:2018yhj,Bagnaschi:2019esc,Murphy:2019qpm}). Since typically no higher-dimensional operators are included,\footnote{See~\cite{Bagnaschi:2017xid} for an EFT calculation including the dominant dimension-six operators.} terms suppressed by the SUSY scale are missed. Therefore at low SUSY scales, the EFT approach is less precise than the fixed-order approach, which includes all suppressed terms up to the order of the calculation. In order to obtain a precise prediction also for intermediary scales, where both a resummation of large logarithms as well as suppressed terms might be relevant, hybrid approaches combining fixed-order and EFT calculations have been developed~\cite{Hahn:2013ria,Bahl:2016brp,Athron:2016fuq,Staub:2017jnp,Bahl:2017aev,Athron:2017fvs,Bahl:2018jom,Bahl:2018ykj,Harlander:2019dge}.

It is, however, crucial for phenomenological studies to not only determine the Higgs-boson mass prediction but also to assess its theoretical uncertainty, $\Delta M_h$, from missing higer-order corrections.\footnote{An additional source of uncertainty is the parametric uncertainty that is induced by the experimental errors of the input parameters entering the calculation, which we will not discuss further in this work (see e.g.\ \cite{Heinemeyer:2003ud}).} The uncertainty of the fixed-order approach has been discussed in~\cite{Degrassi:2002fi,Allanach:2004rh,Allanach:2018fif}, the uncertainty of the EFT approach in~\cite{Hahn:2013ria,Bagnaschi:2014rsa,Vega:2015fna,Bagnaschi:2017xid}. In this work, we discuss the uncertainty of the hybrid approach implemented into the public code \FH~\cite{Heinemeyer:1998yj,Heinemeyer:1998np,Hahn:2009zz, Degrassi:2002fi,Frank:2006yh,Hahn:2013ria,Bahl:2016brp,Bahl:2017aev,Bahl:2018qog}, which combines state-of-the-art fixed-order and EFT calculations.\footnote{The uncertainty of the hybrid approach employed in~\cite{Athron:2016fuq,Staub:2017jnp,Athron:2017fvs} has been estimated in~\cite{Athron:2016fuq,Athron:2017fvs}. Also in~\cite{Harlander:2019dge}, a discussion of the uncertainty in their hybrid approach can be found.} We find that the uncertainty of the combined hybrid approach is lower than or comparable to the individual uncertainties of the fixed-order and EFT calculations in all of the considered parameter space.

This work is structured as follows: In \Sec{sec:02_HYBintro}, we discuss the hybrid approach as implemented in \FH. We then assess the uncertainties of the involved fixed-order calculation in \Sec{sec:03_FOunc} and of the involved EFT calculation in \Sec{sec:04_EFTunc}. The uncertainty estimate of the combined hybrid calculation is discussed in \Sec{sec:05_HYBunc}. We present numerical results in \Sec{sec:06_results} and provide conclusions in \Sec{sec:07_conclusions}.


\section{Hybrid approach for the calculation of the MSSM Higgs boson masses}
\label{sec:02_HYBintro}

In the fixed-order approach, diagrammatic corrections to the Higgs-boson propagators are calculated in the full theory. Taking these self-energy corrections into account, the Higgs-boson masses squared are then determined as the real parts of the propagator poles. The fixed-order calculation implemented in \FH incorporates full one-loop as well as \\\order{\alt\als,\alb\als,\alt^2,\alt\alb,\alb^2} two-loop corrections~\cite{Heinemeyer:1998yj,Heinemeyer:1998np,Degrassi:2001yf, Brignole:2001jy,Brignole:2002bz,Degrassi:2002fi,Dedes:2003km, Heinemeyer:2004xw,Frank:2006yh,Heinemeyer:2007aq,Hahn:2009zz, Hollik:2014wea,Hollik:2014bua,Hollik:2015ema,Hahn:2015gaa} ($\alpha_{b,t} = y_{b,t}^2/(4\pi)$ with $y_{b,t}$ being the bottom and top Yukawa couplings; $\alpha_s = g_3^2/(4\pi)$ with $g_3$ being the strong gauge coupling). Notably, the stop sector of the fixed-order calculation is chosen to be renormalized in the on-shell (OS) scheme (see~\cite{Heinemeyer:2007aq} for more details). For the combination with the EFT calculation in case of \DR stop input parameters, the stop sector can alternatively also be renormalized in the \DR scheme (see \cite{Bahl:2017aev,Bahl:2018qog} for more details).

The parameterization of the top-quark mass is crucial for the Higgs mass prediction, since the leading one-loop corrections are proportional to the top-quark mass to the fourth power. In \FH three different options are available: the OS top-quark mass, the SM \MS top-quark mass (default) and the MSSM \DR top-quark mass. The OS top-quark mass is closely related to the observables used in top-quark mass measurements. Using it, however, leads to (in comparison to the other schemes) larger loop corrections. The leading SM QCD corrections can be absorbed by employing the SM \MS top-quark mass~\cite{Williams:2011bu}. We extract this mass from the OS top-quark mass at the two-loop level including the dominant QCD corrections as well as subleading electroweak corrections using the expressions provided in~\cite{Buttazzo:2013uya} (for testing also the one-loop version is implemented). Also in this scheme large SUSY corrections (i.e., large logarithms of the SUSY scale over the top-quark mass) remain in the calculation. The behaviour in case of a large hierarchy between \msusy and $M_t$ can be improved by defining the top-quark mass in the \DR scheme at the scale \msusy (see e.g.\ discussion in~\cite{Athron:2016fuq}). In \FH, the \DR top-quark mass is calculated taking into account only the leading \order{\als} and \order{\alt} corrections (the electroweak one-loop corrections and also the leading two-loop corrections~\cite{Avdeev:1997sz,Bednyakov:2002sf,Bednyakov:2005kt} are also known but not implemented).
\medskip

In the simplest EFT approach, all SUSY particles are, in contrast to the fixed-order calculation, integrated out at a common scale. Below this SUSY scale, the SM is recovered as EFT. Matching conditions between the SM and the MSSM fix the value of the SM Higgs self-coupling at the matching scale. Renormalization group equations (RGEs) are then used to run the SM couplings down to the electroweak scale at which the Higgs mass is calculated. The EFT calculation implemented in \FH yields a full resummation of leading and next-to-leading logarithms. Moreover, next-to-next-to-leading logarithms (NNLL) proportional to the strong gauge coupling and the top Yukawa coupling are resummed (the electroweak gauge couplings are neglected at the NNLL level). Note, however, that no logarithms proportional to the bottom Yukawa coupling are resummed. In addition to the SM as EFT below the SUSY scales, also the SM with added gauginos and Higgsinos can be used as EFT.\footnote{We do not consider the case of a THDM as EFT below the SUSY scale (see \cite{Bahl:2018jom}) in this paper.} For more details see \cite{Hahn:2013ria,Bahl:2016brp}.

\medskip

The basic idea of the hybrid approach, as implemented in \FH, is to directly add the results of the fixed-order approach -- the renormalized self-energies -- and the results of the EFT approach. For this purpose subtraction terms must be introduced to avoid double-counting of terms contained in the diagrammatic as well as in the EFT result.

First, this concerns large logarithmic contributions. The fixed-order result contains logarithms at the one- and two-loop level. In the EFT result, these logarithms are resummed by RGE evolution of the SM Higgs self-coupling. When adding the two results, we consequently have to subtract the logarithms that would be double-counted.

Second, also non-logarithmic terms which are non-zero in the limit $v/\msusy \to 0$ are contained in both results, where $v$ denotes the vacuum expectation value (which is further specified in the expressions given below) and $\msusy$ the SUSY scale. These non-logarithmic terms are parameterized differently in the fixed-order and the EFT calculation. Mostly OS masses are used for the parameterization of the fixed-order result (see \cite{Frank:2006yh} for a detailed description of the used renormalization scheme). The non-SM terms in the EFT result, which enter via threshold corrections at the SUSY scale, are on the other hand expressed in terms of \MS couplings evaluated at the SUSY scale.

We illustrate this difference by writing down the well-known non-logarithmic one-loop correction originating from the top quark and its scalar superpartners in the two approaches. In the fixed-order calculation, expressed in terms of the SM \MS top quark mass $m_t^\MS$ evaluated at the OS top quark mass $M_t$ (alternatively, the result could also be expressed in terms of the OS top quark mass; the resulting difference would be of two-loop order), it is given by
\begin{align}\label{eq:FOstopnonlog}
(M_h^2)_\text{1L,stop}^\text{FO,non-log} = \frac{12}{(4\pi)^2}
\frac{\left[m_t^\MS(M_t)\right]^4}{v_{G_F}^2} \left[
\left(\frac{X_t}{\msusy}\right)^2 -
\frac{1}{12}\left(\frac{X_t}{\msusy}\right)^4\right] .
\end{align}
Here $v_{G_F}^2 = 1/(2\sqrt{2}G_F)$ (with the Fermi constant $G_F$), and $X_t$ is the stop mixing parameter which when multiplied with the top quark mass constitutes the off-diagonal entry of the stop mass matrix. For simplicity, we also assumed the stop soft-SUSY breaking mass parameters to be equal to $\msusy$ and we neglected terms suppressed by $v_{G_F}/\msusy$.

In the EFT approach, the corresponding terms enter as a threshold correction at the matching scale $\msusy$,
\begin{align}\label{eq:EFTstopnonlog}
(M_h^2)_\text{1L,stop}^\text{EFT,non-log} &= \frac{12}{(4\pi)^2} y_t^4(\msusy)
\left[v^\MS(M_t)\right]^2 \left[ \left(\frac{X_t}{\msusy}\right)^2 - \frac{1}{12}\left(\frac{X_t}{\msusy}\right)^4\right] =\nonumber\\
&= \frac{12}{(4\pi)^2} y_t^4(M_t)
\left[v^\MS(M_t)\right]^2 \left[ \left(\frac{X_t}{\msusy}\right)^2 - \frac{1}{12}\left(\frac{X_t}{\msusy}\right)^4\right] + \text{logs},
\end{align}
where $v^\MS$ is the SM \MS vev, entering at the electroweak scale, and $y_t = m_t^\MS/v^\MS$ is the SM top Yukawa coupling in the \MS scheme.

Since the threshold corrections obtained in the EFT approach, as given in the first line of \Eq{eq:EFTstopnonlog}, are needed to correctly take into account higher-order logarithmic contributions, they need to appear in the same form also in the hybrid result. As illustrated in the second line of \Eq{eq:EFTstopnonlog}, the parameterization of the fixed-order result of \Eq{eq:FOstopnonlog} would differ from the parameterization of the EFT result by higher-order logarithmic terms. Thus, the subtraction terms in the hybrid approach need to be chosen such that the non-logarithmic terms of the fixed-order calculation (see \Eq{eq:FOstopnonlog}) are cancelled in the limit $v/\msusy \to 0$.\footnote{In previous \FH versions, the non-logarithmic terms of the EFT result reexpressed in terms of \MS couplings at the scale $M_t$ were subtracted instead. This leads to small differences, \order{100\mev}, in the final value for $M_h$ (for more details see \cite{Bahl:2017aev}). For $\msusy < 1\tev$, larger shifts are possible (see \Sec{sec:06_results}).} The only exception are two-loop contributions induced by fixing the parameters of the stop sector in the OS scheme. These appear only in the fixed-order calculation and therefore are not subtracted in the hybrid result.

In order to obtain the hybrid result particular care is necessary if the input parameters of the fixed-order and the EFT calculation are not defined in the same renormalization scheme. Especially relevant is the definition of the stop mixing parameter $X_t$, since it has a large impact on the Higgs mass prediction. In the EFT calculation, $X_t$ is defined in the \DR scheme. The fixed-order calculation allows one to define $X_t$ either in the OS or the \DR scheme (the latter is only employed in our approach when embedding the fixed-order contribution into the hybrid result). If the OS scheme is employed, we need to convert $X_t$ from the OS to the \DR scheme before using it as input for the EFT calculation. It is sufficient for this conversion to take only large one-loop logarithms into account, since only these terms are needed to reproduce the logarithms contained in the fixed-order approach \cite{Bahl:2016brp,Bahl:2017aev},
\begin{align}\label{eq:Xtconv}
X_t^\DR(\msusy) = X_t^\OS \left\{1 + \left[\frac{\alpha_s}{\pi} - \frac{3\alpha_t}{16\pi}\left(1 - \left(\frac{X_t}{\msusy}\right)^2\right)\right]\ln\frac{\msusy^2}{M_t^2}\right\}.
\end{align}
For the other input parameters no conversion is necessary since no large logarithms appear in the conversion formulas.\footnote{Since the bottom Yukawa coupling is neglected in our EFT calculation up to now, the sbottom mixing parameter, $X_b$, whose OS to \DR conversion involves large logarithms, does not enter the EFT calculation.} As mentioned above, in the case of OS input parameters not all non-logarithmic terms contained in the fixed-order calculation that are unsuppressed in the limit of large \msusy should be subtracted. Those terms originating from OS stop counterterms are not generated in the EFT calculation due to the log-only conversion of the stop parameters. Therefore, no subtraction of these terms is needed.


\section{Uncertainty of the fixed-order calculation}
\label{sec:03_FOunc}

The uncertainty of the fixed-order calculation implemented in \FH has been discussed previously in~\cite{Degrassi:2002fi}. The authors proposed two different methods to assess the uncertainty: varying the renormalization scale in the interval $[M_t/2,2M_t]$ and using different renormalization schemes for the top-quark mass. Here, we  largely follow this prescription. Moreover, deactivating the resummation of the bottom-Yukawa coupling for large $\tan\beta$ (see~\cite{Carena:1999py} for more details) has been employed as an additional method to estimate the uncertainty

We, however, do not include a variation of the renormalization scale into our uncertainty estimate, since, as shown in~\cite{Bahl:2018ykj}, the dependence of the fixed-order calculation on the renormalization scale is completely dominated by terms induced through the Higgs pole determination, which would cancel order-by-order in a more complete calculation. The uncertainty of terms induced by the Higgs pole determination (i.e.\ by the momentum dependence of the SM-like contributions to the Higgs self-energies) is already assessed by switching between the different renormalization schemes for the top-quark mass.

Instead of varying the renormalization scale, we consider higher-order QCD corrections as an additional source of uncertainty (see also the discussion in \cite{Drechsel:2016htw}). This concerns especially three-loop leading logarithms, whose size is not estimated by using different renormalization schemes for the top-quark mass.\footnote{Additional \order{\alt\als^2} fixed-order corrections presented in~\cite{Harlander:2008ju,Kant:2010tf,Harlander:2017kuc} are not yet included in the fixed-order calculation implemented in \FH.} We estimate the size of those uncertainties by multiplying the two-loop \order{\alt\als} correction with $\als/(4\pi)\ln(\msusy^2/M_t^2)$ (from now on, we define $\msusy$ as the geometric mean of the stop soft-SUSY breaking masses). Thus, in order to estimate those uncertainties in the fixed-order calculation we replace $\als$ by $\als\left[1 \pm \als/(4\pi)\ln(\msusy^2/M_t^2)\right]$.

In principle the uncertainty associated with those three-loop leading logarithms could also be estimated by evaluating the strong gauge coupling $\als$ at a different scale, for instance $\msusy$, instead of the default choice $M_t$. However, in this case also the top-quark mass should be evaluated at $\msusy$ in order to avoid artificially large contributions arising from the different scale choices. Since the fixed-order calculation implemented in \FH does not include this parametrization of the \MS top-quark mass, we keep $\als$ at the scale $M_t$ and perform the uncertainty estimate as described above.

Summing up, we estimate the uncertainty of the fixed-order calculation by:
\begin{itemize}
\item switching between different parametrizations of the top-quark mass (OS top-quark mass and SM~\MS top-quark mass at the one- and two-loop level, evaluated at the scale $M_t$),
\item switching on and off the resummation of the bottom-Yukawa coupling for large $\tan\beta$,
\item replacing $\als$ by $\als\left[1 \pm \als/(4\pi)\ln(\msusy^2/M_t^2)\right]$.
\end{itemize}
The absolute values of the respective shifts obtained for each observable are added linearly.


\section{Uncertainty of the EFT calculation}
\label{sec:04_EFTunc}

Following~\cite{Bagnaschi:2014rsa,Vega:2015fna}, there are three sources of uncertainty for a pure EFT calculation,
\begin{itemize}
\item High-scale uncertainty: uncertainty associated with higher-order threshold corrections,
\item Low-scale uncertainty: uncertainty in the extraction of the low-energy couplings used as input for the RGE running as well as uncertainty in the determination of the Higgs pole mass at the low scale,
\item Uncertainty from \order{v/\msusy} terms: uncertainty associated with not-included terms suppressed by the SUSY scale.
\end{itemize}
In principle, there is also an uncertainty associated with unknown higher-order RGE running effects. This is, however, expected to be a negligible source of uncertainty, since already the known three-loop RGE running effects only have a very small impact on the mass of the SM-like Higgs boson (see e.g.~\cite{Lee:2015uza}).

Following the prescriptions in~\cite{Bagnaschi:2014rsa,Vega:2015fna,Allanach:2018fif,Harlander:2019dge}, we estimate the individual uncertainties as follows,
\begin{itemize}
\item High-scale uncertainty
\begin{itemize}
\item varying the high-energy matching scale $Q_\text{match}$ between the full MSSM and the low-energy EFT in the interval $[\msusy/2,2\msusy]$,
\item reparametrizing the threshold corrections between the low-energy EFT and the full MSSM in terms of the MSSM top Yukawa coupling (by default they are expressed in terms of the SM top Yukawa coupling).
\end{itemize}
Since we take into account the full one-loop threshold correction for the Higgs self-coupling including electroweak contributions, this procedure also assesses the possible size of two-loop electroweak terms.
\item Low-scale uncertainty:
\begin{itemize}
\item switching between an extraction of the SM \MS top Yukawa coupling from the OS top-quark mass at the two- and three-loop level (by default the two-loop SM top Yukawa coupling is used),
\item finding the Higgs pole mass employing either the OS top-quark mass or the SM \MS top-quark mass, evaluated at the scale $M_t$, in the SM Higgs self-energy,
\end{itemize}
\item Uncertainty from \order{v/\msusy} terms:
\begin{itemize}
\item multiplying the one-loop threshold correction for the Higgs self-coupling by\\ $2v^2/\msusy^2$ and adding it to or subtracting it from the unsuppressed threshold correction.
\end{itemize}
\end{itemize}
As in \Sec{sec:03_FOunc}, all individual contributions are added linearly.

We adapt the uncertainty estimate of the SM corrections in the EFT below the SUSY scale with respect to~\cite{Bagnaschi:2014rsa,Vega:2015fna,Allanach:2018fif,Harlander:2019dge}. In~\cite{Bagnaschi:2014rsa,Vega:2015fna}, a fixed numerical estimate of $\sim 0.15\gev$,  obtained in \cite{Buttazzo:2013uya}, was used. In~\cite{Allanach:2018fif}, the uncertainty associated with SM corrections was assessed by varying the low-energy scale at which the poles are determined in the interval $[M_t/2,2M_t]$. Using instead different renormalization schemes for the top-quark mass allows for an easier combination of the fixed-order and the EFT uncertainty estimates for the hybrid calculation.

In order to obtain an alternative independent cross-check for the high-scale uncertainty, we implemented the \order{\alt\als^2} threshold correction presented in \cite{Harlander:2018yhj} by linking the publicly available code \texttt{Himalaya} \cite{Harlander:2017kuc,Harlander:2018yhj} to \FH. As discussed in \cite{Harlander:2018yhj}, the result for the \order{\alt\als^2} threshold correction itself is affected by an uncertainty since the result is based upon an expansion of three-loop diagrams. For our cross-check, we add the central value for the \order{\alt\als^2} threshold correction plus or minus the uncertainty estimate of this value provided by \texttt{Himalaya} to the existing one- and two-loop threshold corrections. We then calculate the shifts in $M_h$ and compare the larger shift to our high-scale uncertainty estimate obtained as outlined above.


\section{Uncertainty of the combined hybrid calculation}
\label{sec:05_HYBunc}

The hybrid approach profits from the advantages of both, the fixed-order and the EFT approach. Therefore, the uncertainty of the hybrid calculation can be expected, depending on the considered parameter region, to be either below or at most as big as the separate uncertainties of the involved fixed-order and EFT calculations.

The major source of uncertainty in the fixed-order approach are higher-order logarithmic contributions. The EFT result resums all large logarithmic contributions contained in the fixed-order approach if for every large hierarchy between different mass scales a corresponding EFT prescription is implemented. Therefore, the logarithmic uncertainty of the hybrid approach should correspond to the logarithmic uncertainty of the EFT result. Accordingly, for estimating the uncertainty associated with higher-order logarithmic contributions in the hybrid approach we employ the same estimate as for the pure EFT calculation.

The major source of uncertainty in the EFT approach for (some) low SUSY scales are terms of \order{v/\msusy}. In the fixed-order approach, all terms which would be suppressed for high SUSY scale are fully included at the one-loop level and at the two-loop level in the limit of vanishing electroweak gauge couplings. Therefore, the estimate of the uncertainty of the combined hybrid calculation should include an assessment of terms of \order{v/\msusy} only beyond the one- and dominant two-loop level. We estimate the uncertainty of those terms with the same method that is also used to assess the uncertainty of suppressed terms in the fixed-order approach.

An important source of uncertainty for the hybrid result are unsuppressed higher-order non-logarithmic terms, which do not explicitly depend on $\msusy$ if $X_t/\msusy$ is kept constant. The unsuppressed non-logarithmic terms are parameterized differently in the fixed-order and the EFT calculation. As discussed in \Sec{sec:02_HYBintro}, in order to correctly take into account higher-order logarithmic contributions the unsuppressed non-logarithmic terms of the fixed-order calculation have to be subtracted in the combined hybrid result. This procedure is crucial not only for the prediction of $M_h$ but also for the estimate of the associated theoretical uncertainty.

Concerning the latter, this prescription has important implications on the estimated size of the uncertainties. If the uncertainty associated with unsuppressed non-logarithmic terms were estimated based on the expressions appearing in the fixed-order calculation (see \Eq{eq:FOstopnonlog}), the resulting estimate would stay constant when varying \msusy (but keeping $X_t/\msusy$ constant) since the coefficient in front of the square bracket of \Eq{eq:FOstopnonlog} does not explicitly depend on \msusy. In contrast, the estimate of the theoretical uncertainty associated with the non-logarithmic terms as appearing in the EFT calculation (see \Eq{eq:EFTstopnonlog}) is reduced with rising \msusy reflecting the decrease of the top Yukawa coupling evaluated at \msusy, $y_t(\msusy)$, with rising \msusy. As indicated in \Eq{eq:EFTstopnonlog}, the different parameterizations of the coefficient of the non-logarithmic terms amount to a difference in logarithmic higher-order terms. The resulting estimate of the theoretical uncertainty of the non-logarithmic terms following the described prescription yields, as we will show below, the expected behaviour of a decrease with increasing \msusy, since also the numerical impact of the non-logarithmic terms on the prediction of $M_h$ itself decreases with rising \msusy.

The situation is different for the terms that are suppressed by powers of $v/\msusy$, since these contributions only appear in the fixed-order result. As explained above, we assess the uncertainty associated with these suppressed terms analogously to the fixed-order approach.

We now turn to the uncertainties that are associated with SM-like corrections. These uncertainties are estimated in the fixed-order and the EFT approach in a similar way by employing different renormalization schemes for the top-quark mass.

An additional source of uncertainty arises if the OS scheme is used for the definition of the input parameters of the stop sector in the fixed-order part of the calculation. First, the conversion of $X_t$ between the OS and the \DR scheme (taking into account only logarithmic contributions, see \Eq{eq:Xtconv}) is affected by higher-order uncertainties. We assess these by replacing $\als$ entering \Eq{eq:Xtconv} by $\als\left[1 \pm \als/(4\pi)\ln(\msusy^2/M_t^2)\right]$ and by switching between different definitions of the top-quark mass, which enters \Eq{eq:Xtconv} via the top Yukawa coupling. Second, the OS renormalization of the stop parameters induces non-logarithmic contributions to the Higgs self-energies. Correspondingly, there is an associated uncertainty induced by higher-order contributions to the OS stop counterterms which is not accounted for by the estimate of higher-order non-logarithmic contributions in the EFT part. We parameterize the known two-loop contributions in terms of low-energy couplings (as in the fixed-order calculation). Therefore, also this uncertainty can be estimated by replacing $\als$ by $\als\left[1 \pm \als/(4\pi)\ln(\msusy^2/M_t^2)\right]$ and switching between different definitions of the top-quark mass.

It should be noted in this context that a Higgs-mass prediction in terms of \DR input parameters (the same applies to all other parameters that cannot directly be associated with physical observables) is only a part of a physically meaningful prediction relating physical observables to each other. If \DR parameters are employed in the Higgs-mass prediction it is therefore unavoidable to invoke well-defined prescriptions relating the \DR parameters to physical observables. The evaluation of those relations (depending on the level of sophistication with which it is carried out) introduces an additional theoretical uncertainty that needs to be taken into account in the overall relation between physical observables.

In total, our estimate of the remaining theoretical uncertainties in the hybrid approach consists of the following elements:
\begin{itemize}
\item High-scale uncertainty (estimated in the EFT part):
\begin{itemize}
\item varying the high-energy matching scale $Q_\text{match}$ between the full MSSM and the low-energy EFT in the interval $[\msusy/2,2\msusy]$,
\item reparametrizing the threshold corrections between the low-energy EFT and the full MSSM in terms of the MSSM top Yukawa coupling (by default they are expressed in terms of the SM top Yukawa coupling).
\end{itemize}
\item Low-scale uncertainty (estimated in the fixed-order part):
\begin{itemize}
\item switching between an extraction of the SM \MS top Yukawa coupling from the OS top mass at the two- and three-loop level (by default the two-loop SM top Yukawa coupling is used),
\item finding the Higgs pole mass employing either the OS top-quark mass or the SM \MS top-quark mass, evaluated at the scale $M_t$, at the two-loop level in the Higgs self-energy.
\end{itemize}
\item Uncertainty from \order{v/\msusy} terms (estimated in the fixed-order part):
\begin{itemize}
\item switching between different parametrizations of the top-quark mass (OS
top-quark mass and SM \MS\ top-quark mass,
evaluated at the scale $M_t$,
at the NNLO level),
\item switching on and off the resummation of the bottom-Yukawa coupling for large $\tan\beta$,
\item replacing $\als$ by $\als\left[1 \pm \als/(4\pi)\ln(\msusy^2/M_t^2)\right]$.
\end{itemize}
\end{itemize}
If the stop sector is renormalized using the OS scheme, there are two additional contributions (as discussed above, if \DR parameters are employed, additional theoretical uncertainties occur in the relation between the predicted Higgs-boson mass and other physical observables):
\begin{itemize}
\item Uncertainty in the calculation of $X_t^\DR$ needed in the EFT calculation:
\begin{itemize}
\item switching between different parametrizations of the top-quark mass (OS top-quark mass and SM \MS\ top-quark mass, evaluated at the scale $M_t$, at the NNLO level),
\item replacing $\als$ by $\als\left[1 \pm \als/(4\pi)\ln(\msusy^2/M_t^2)\right]$.
\end{itemize}
\item Uncertainty from higher-order non-logarithmic contributions to the stop counterterms:
\begin{itemize}
\item switching between different parametrizations of the top-quark mass (OS top-quark mass and SM \MS\ top-quark mass, evaluated at the scale $M_t$, at the NNLO level),
\item replacing $\als$ by $\als\left[1 \pm \als/(4\pi)\ln(\msusy^2/M_t^2)\right]$.
\end{itemize}
\end{itemize}
In case of complex input parameters, there is an additional source of uncertainty (and also in this case the uncertainty in relating those parameters to physical observables has to be taken into account). Since the EFT calculation and a part of the two-loop contributions of the fixed-order result are only known for the case of real input parameters up to now, an interpolation method is used for the evaluation of complex input parameters (see \cite{Bahl:2018qog} for more details). This uncertainty is not assessed by the prescription presented above. The same is true for the case of non-minimal flavour violation. Non-minimal flavour violation is only taken into account at the one-loop level in the diagrammatic calculation~\cite{AranaCatania:2011ak,Heinemeyer:2004by,Gomez:2014uha}.


\section{Numerical results}
\label{sec:06_results}

In this Section, we investigate numerically the different sources of uncertainty. As a first step, we focus on a simple single-scale scenario, in which all non-SM masses are chosen to be equal to $\msusy$ (including $M_A$ and the Higgsino as well as the gaugino mass parameters). All trilinear soft-SUSY breaking couplings are set to zero apart from the stop trilinear coupling, which is fixed by setting the stop mixing parameter $X_t$. For most results, we set $\tan\beta=20$. Below, we also briefly investigate scenarios with more than one relevant mass scale. If not stated otherwise, the fixed-order calculation employing the SM \MS top-quark mass is meant when we refer to the fixed-order calculation in general.

\medskip

\begin{figure}\centering
\begin{minipage}{.48\textwidth}\centering
\includegraphics[width=\textwidth]{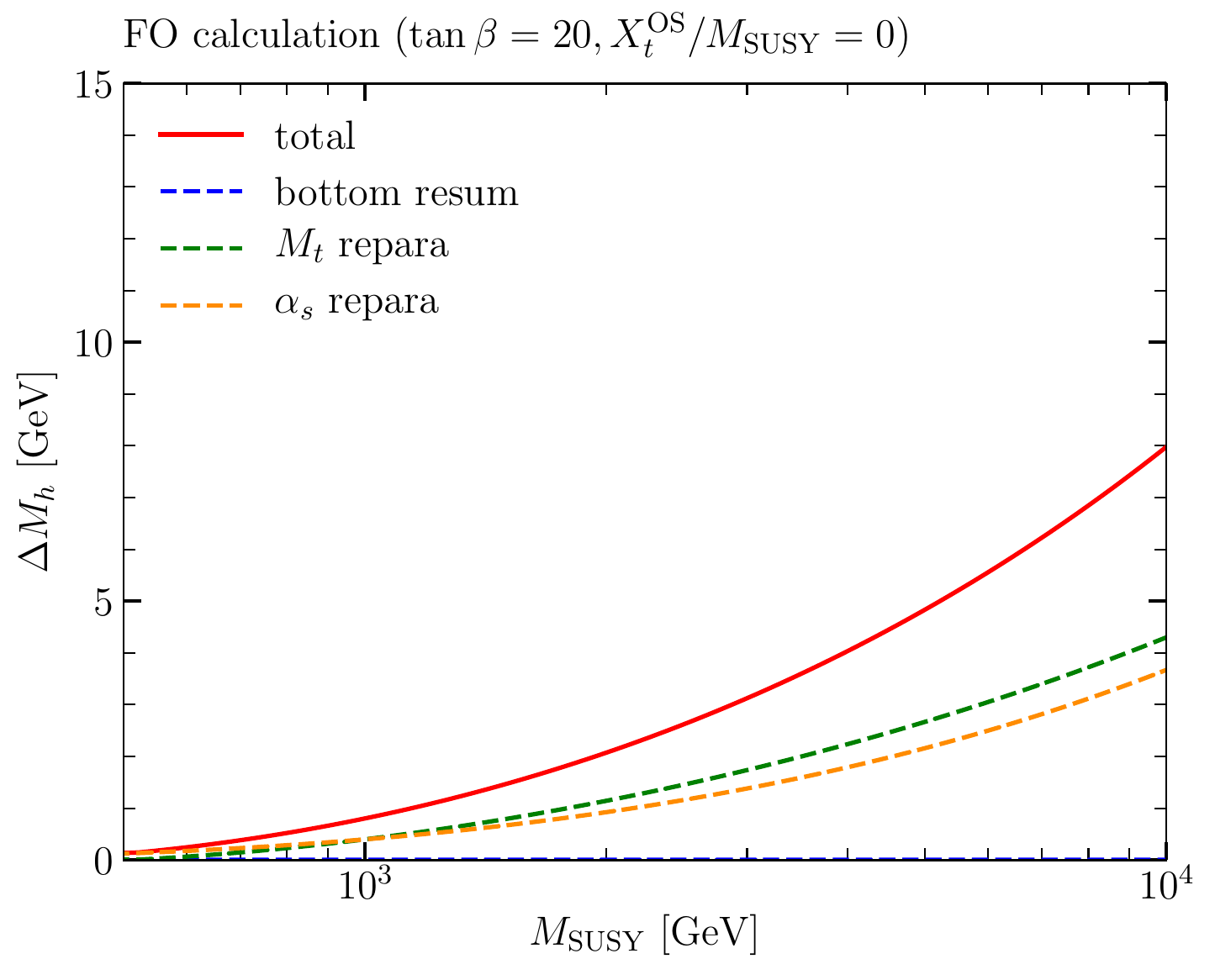}
\end{minipage}
\begin{minipage}{.48\textwidth}\centering
\includegraphics[width=\textwidth]{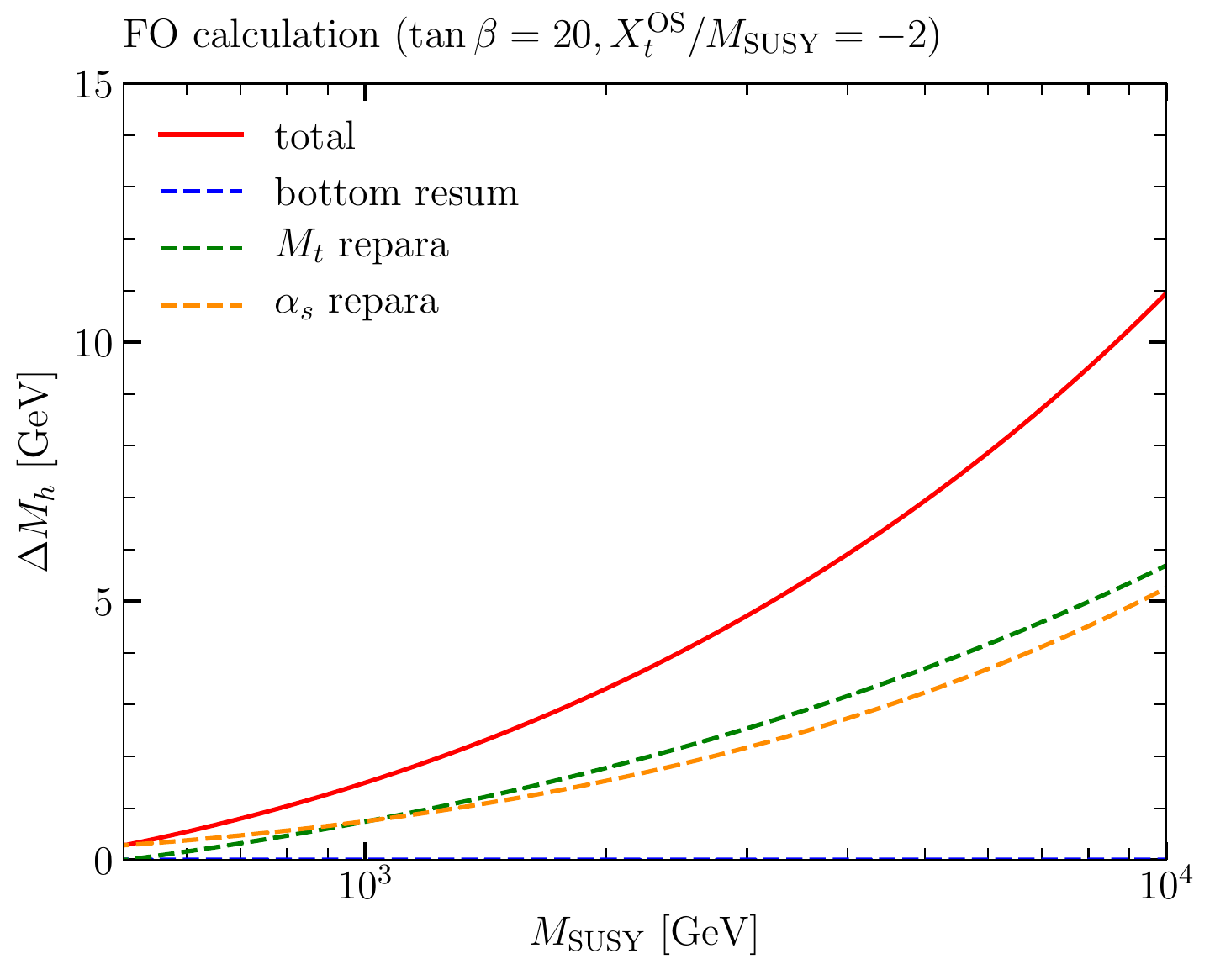}
\end{minipage}
\caption{Uncertainty estimate of the fixed-order calculation with OS renormalized stop sector as a function of $\msusy$ for $\tan\beta = 20$. The stop mixing parameter is chosen to be $X_t^\OS/\msusy = 0$ (left) and $X_t^\OS/\msusy = -2$ (right).}
\label{fig:HSsrcFO_OS}
\end{figure}

First, we investigate the uncertainty of the pure fixed-order calculation in \Fig{fig:HSsrcFO_OS}. As stressed before, as default the OS scheme is used for the renormalization of the stop sector. The various components of the uncertainty estimate are shown as a function of $\msusy$ for vanishing stop mixing (left) and $\xtOS=-2$ (right).\footnote{We choose $X_t$ negative in order to prevent the lighter stop from getting tachyonic for low SUSY scales. Moreover, this choice allows for an easier comparison with results of \cite{Allanach:2018fif}.} As expected, the total uncertainty estimate (red) increases with rising $\msusy$. For vanishing stop mixing the estimated uncertainty is below $\sim 1 \gev$ for $\msusy\lesssim 1\tev$, while it rises to about $5 \gev$ for $\msusy = 5\tev$. This behaviour reflects the increasing importance of higher-order logarithms which are not resummed in the fixed-order approach. Concerning the individual sources of uncertainty, we see that replacing $\als$ by $\als\left[1 \pm \als/(4\pi)\ln(\msusy^2/M_t^2)\right]$ (orange dashed) and switching between the OS top-quark mass and the SM \MS top-quark mass $m_t^\MS(M_t)$ (green dashed) contribute approximately by the same amount to the total uncertainty estimate. The uncertainty associated with switching on and off the resummation of the bottom Yukawa coupling (blue dashed) is negligible for $\tan\beta = 20$ over the whole considered $\msusy$ range.\footnote{The uncertainty associated with the resummation of the bottom Yukawa coupling yields a sizeable contribution for large $\tan\beta$. It can be further enhanced if $\mu$ is negative. The numerical effect of resumming the bottom Yukawa copuling is discussed in detail in~\cite{Brignole:2002bz,Frank:2013hba,Bahl:2020xxx}.} For $\xtOS=-2$ (right plot), we see that the overall behaviour is similar to the case of vanishing stop mixing but with somewhat larger values of the estimated uncertainties. The size of the uncertainty estimate reaches values of up to $\sim 1.5 \gev$ for $\msusy \sim 1\tev$ and $\sim 11 \gev$ for $\msusy \sim 10\tev$.

\begin{figure}\centering
\begin{minipage}{.48\textwidth}\centering
\includegraphics[width=\textwidth]{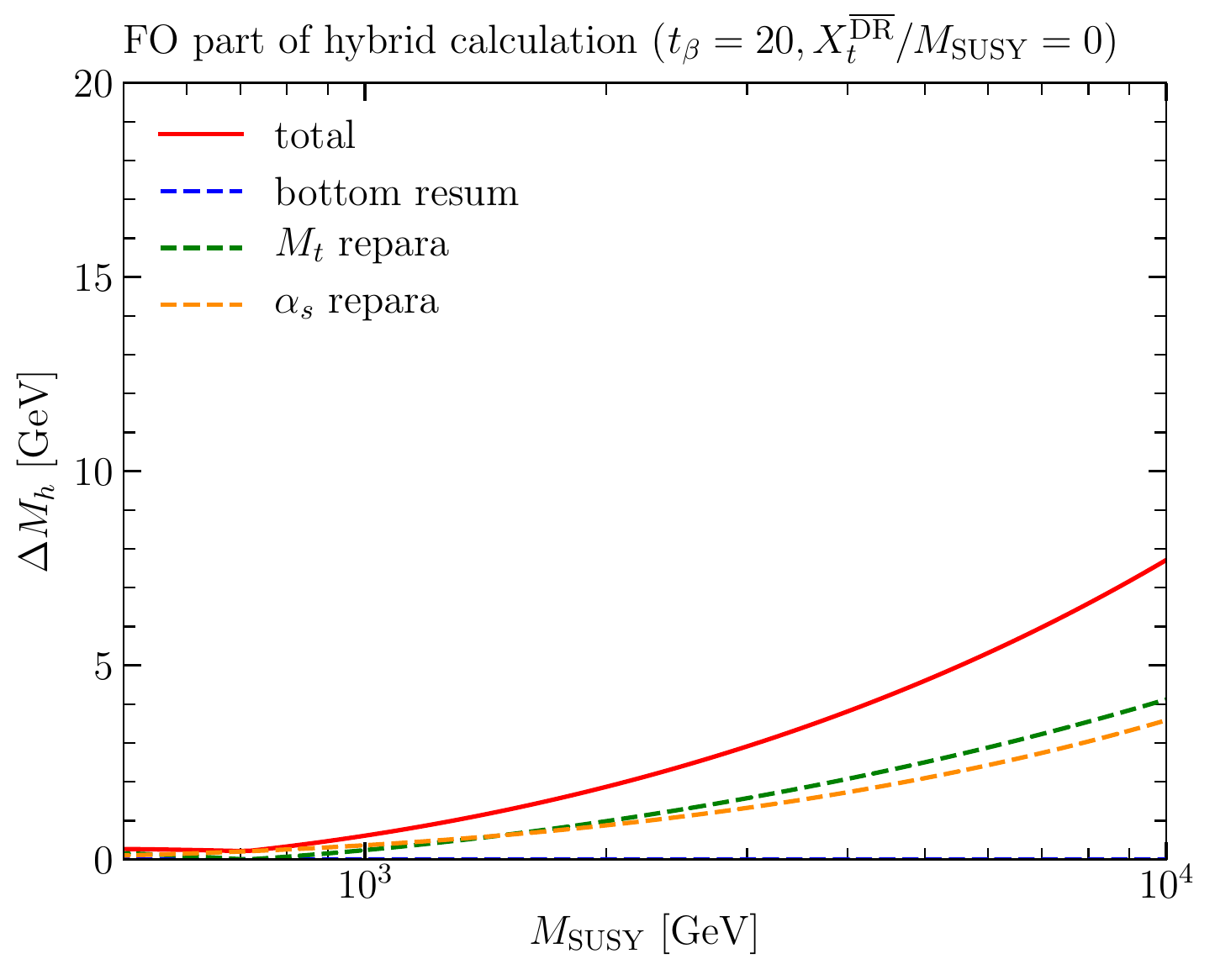}
\end{minipage}
\begin{minipage}{.48\textwidth}\centering
\includegraphics[width=\textwidth]{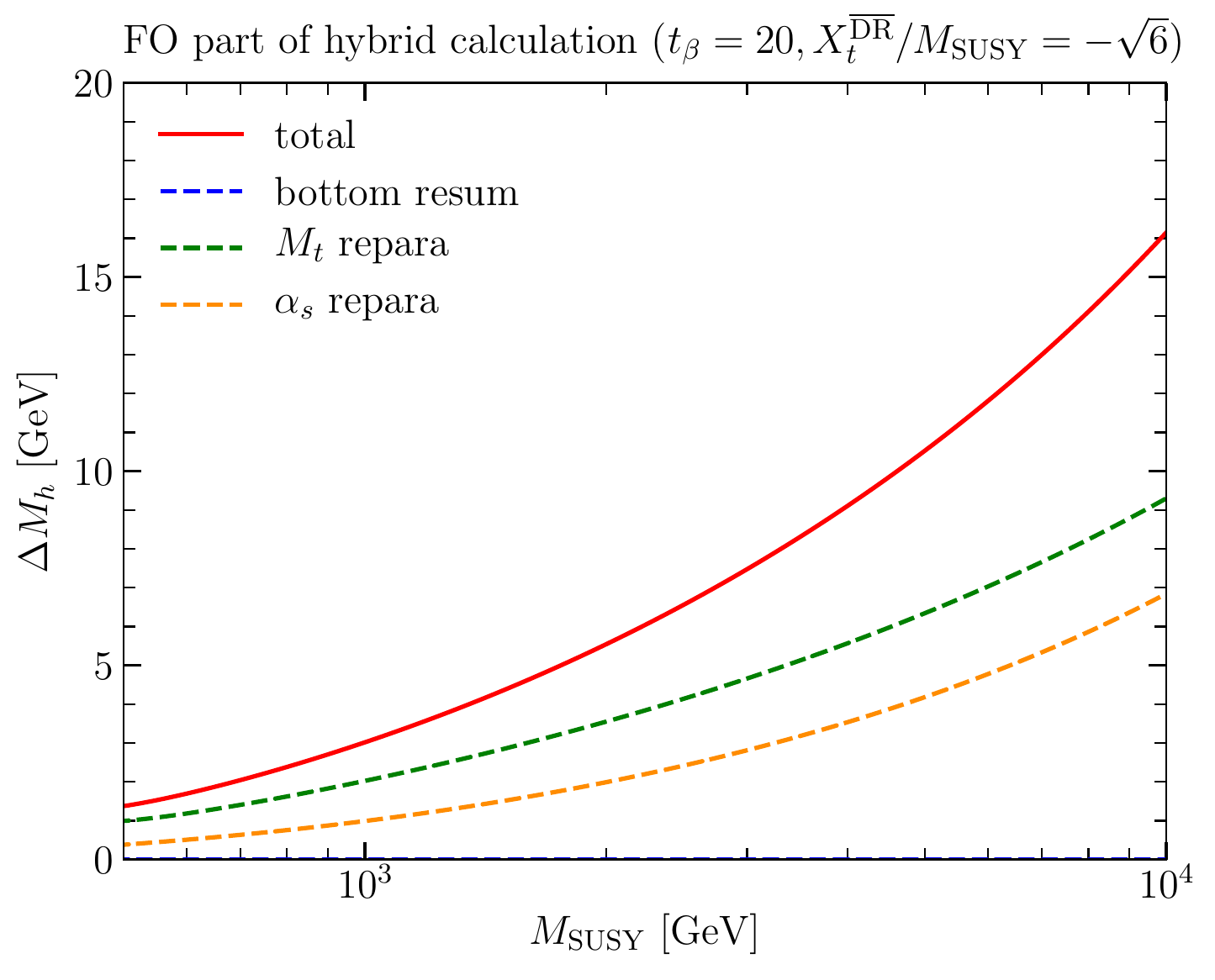}
\end{minipage}
\caption{Uncertainty estimate of the fixed-order part of the hybrid calculation with \DR renormalized stop sector as a function of $\msusy$ for $\tan\beta = 20$. The stop mixing parameter is chosen to be $X_t^\DR/\msusy = 0$ (left) and ${X_t^\DR/\msusy = -\sqrt{6}}$ (right).}
\label{fig:HSsrcFO_DR}
\end{figure}

In \Fig{fig:HSsrcFO_DR}, we show the uncertainty estimate for the fixed-order part of the hybrid calculation (including the logarithmic contributions which are subtracted when this part is combined with the EFT calculation) where the \DR scheme is used for the renormalization of the stop sector. In this case the stop parameters are renormalized in the \DR scheme at the scale \msusy while for the top mass the SM \MS mass at the scale $M_t$ is used. While for vanishing stop mixing (left plot), the uncertainty estimate is very similar to the case of an OS renormalized stop sector (note that the scale of this plot is different than in \Fig{fig:HSsrcFO_OS}), the uncertainty estimate rises to values of $\sim 16\gev$ for $\xtDR = -\sqrt{6}$ and $\msusy\sim 10\tev$ (right plot). Even though the right plot of \Fig{fig:HSsrcFO_OS} and the right plot of \Fig{fig:HSsrcFO_DR} cannot be compared directly (because of the different parameterization of \msusy and $X_t$), this enlarged uncertainty estimate indicates that the fixed-order part of the hybrid calculation implemented in \FH is perturbatively better behaved in the case of the stop sector being renormalized using the OS scheme (which is the default setting of \FH and corresponds to the method that was used for obtaining the implemented results). It should be stressed that the result displayed in \Fig{fig:HSsrcFO_DR} is just a technical ingredient of the full hybrid result and should not be regarded as a valid fixed-order prediction on its own. The occurrence of rather large uncertainties already for relatively low values of $\msusy$ (the estimated uncertainty amounts to $\sim 3 \gev$ for $\msusy \sim 1\tev$) can be traced to the different scale choices for the top-quark mass and the stop parameters, see e.g.\ the discussion of a pure \DR fixed-order calculation in \cite{Allanach:2018fif} and also the discussion of \Fig{fig:HScomp_DR} below. The combination of the fixed-order part shown in \Fig{fig:HSsrcFO_DR} with the EFT calculation leads to a large reduction of the associated uncertainty estimate (see the discussion of \Fig{fig:HSsrcHybrid} below).

\medskip

\begin{figure}\centering
\begin{minipage}{.48\textwidth}\centering
\includegraphics[width=\textwidth]{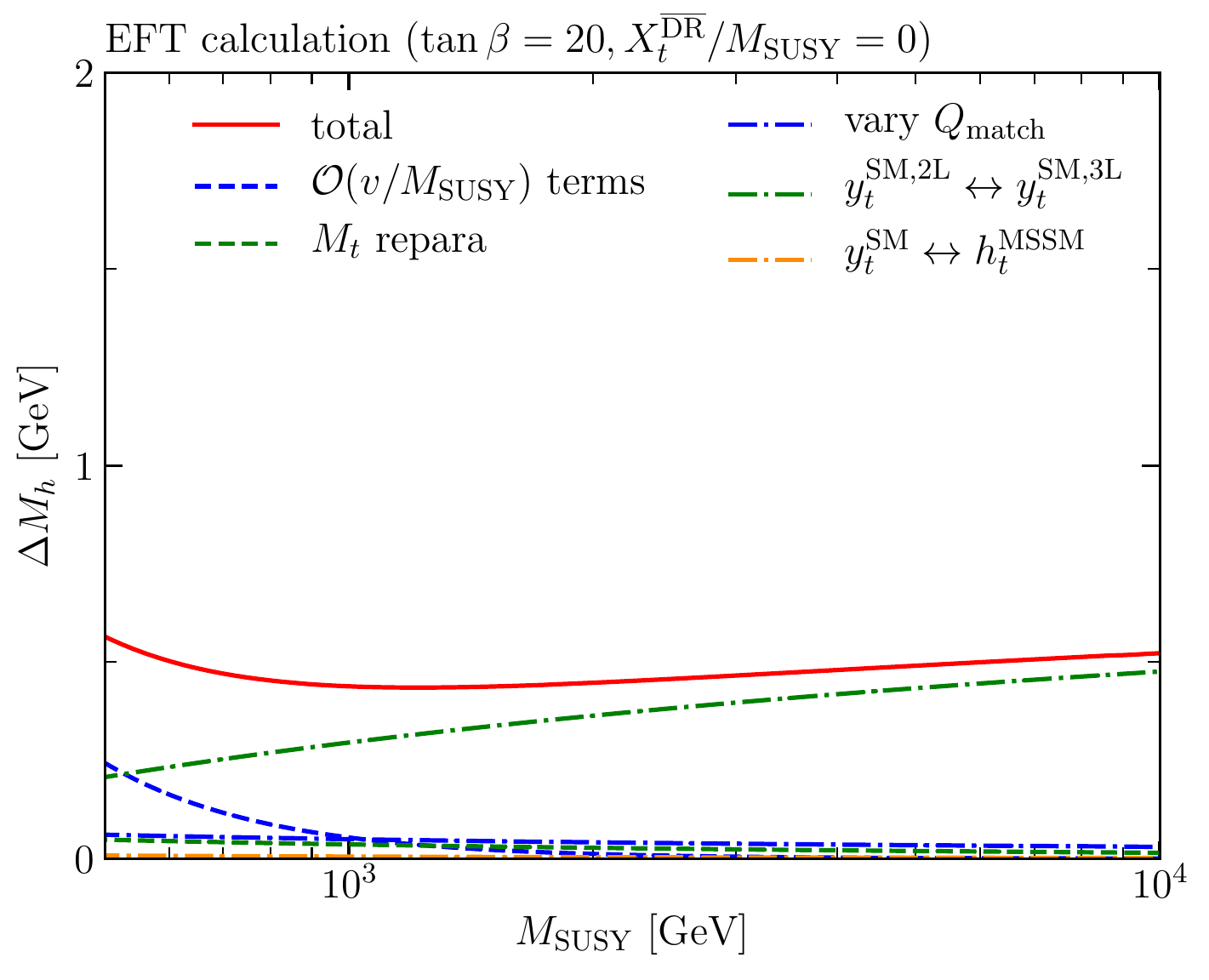}
\end{minipage}
\begin{minipage}{.48\textwidth}\centering
\includegraphics[width=\textwidth]{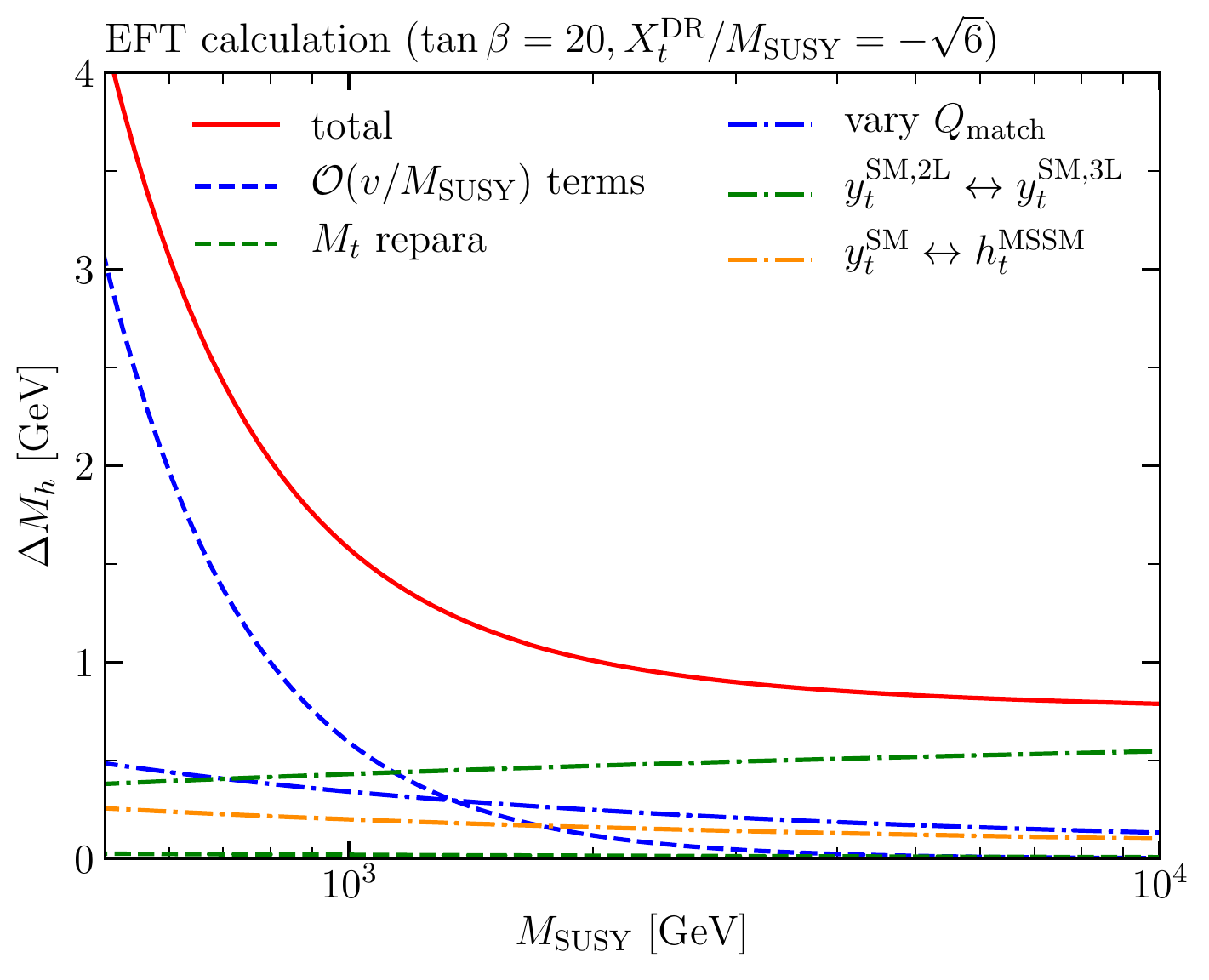}
\end{minipage}
\caption{Left: Uncertainty estimate of the EFT calculation as a function of $\msusy$ for $\tan\beta = 20$ and $X_t^\DR/\msusy = 0$. Right: Same as left plot, but ${X_t^\DR/\msusy = -\sqrt{6}}$.}
\label{fig:HSsrcEFT}
\end{figure}

Next, we assess the uncertainty of the pure EFT calculation in \Fig{fig:HSsrcEFT} for the case of \DR renormalized stop sector parameters. In the left plot, the case of $X_t^\DR/\msusy = 0$ is depicted. The total uncertainty estimate (red) stays almost constant (at $\sim 0.7\gev$) over the displayed range of \msusy. The slight increase for $\msusy\lesssim 1\tev$ reflects the increasing importance of terms suppressed by the SUSY scale (blue dashed). For larger $\msusy$ values the slight upwards trend is caused by the uncertainty that is estimated by extracting the SM \MS top Yukawa coupling at the three-loop level instead of the two-loop level (green dot-dashed). The uncertainties associated with the variation of the matching scale (blue dot-dashed) and the reparametrization of the threshold corrections in terms of the MSSM top Yukawa coupling (orange dot-dashed) play only a minor role. Their decreasing behaviour reflects the decrease of the top Yukawa coupling and the strong gauge coupling with rising $\msusy$. The estimate of the uncertainty of the SM-like corrections obtained by using different parameterizations for the top quark mass (green dashed) yields values of $\sim 0.1 \gev$.

In the right plot of \Fig{fig:HSsrcEFT}, we display the results for $X_t^\DR/\msusy = -\sqrt{6}$. For this value of $X_t^\DR$, the estimate of the impact of the terms that are suppressed by powers of $v/\msusy$ yields much larger values of up to $\sim 4 \gev$ for $\msusy \sim 500\gev$. Also the size of the uncertainty associated with a variation of the matching scale is enlarged in comparison to the case of vanishing stop mixing. For $\msusy\gtrsim 2\tev$ the overall behaviour tends to be similar to the case of vanishing stop mixing. The total size of the uncertainty estimate is $\sim 1\gev$ in this parameter region.

These findings are in good agreement with the EFT results presented in \cite{Bagnaschi:2014rsa,Vega:2015fna,Allanach:2018fif}. In there, only the estimate of the uncertainties of the SM-like corrections was found to be slightly higher (by $\sim 0.1\gev$).

\medskip

\begin{figure}\centering
\begin{minipage}{.48\textwidth}\centering
\includegraphics[width=\textwidth]{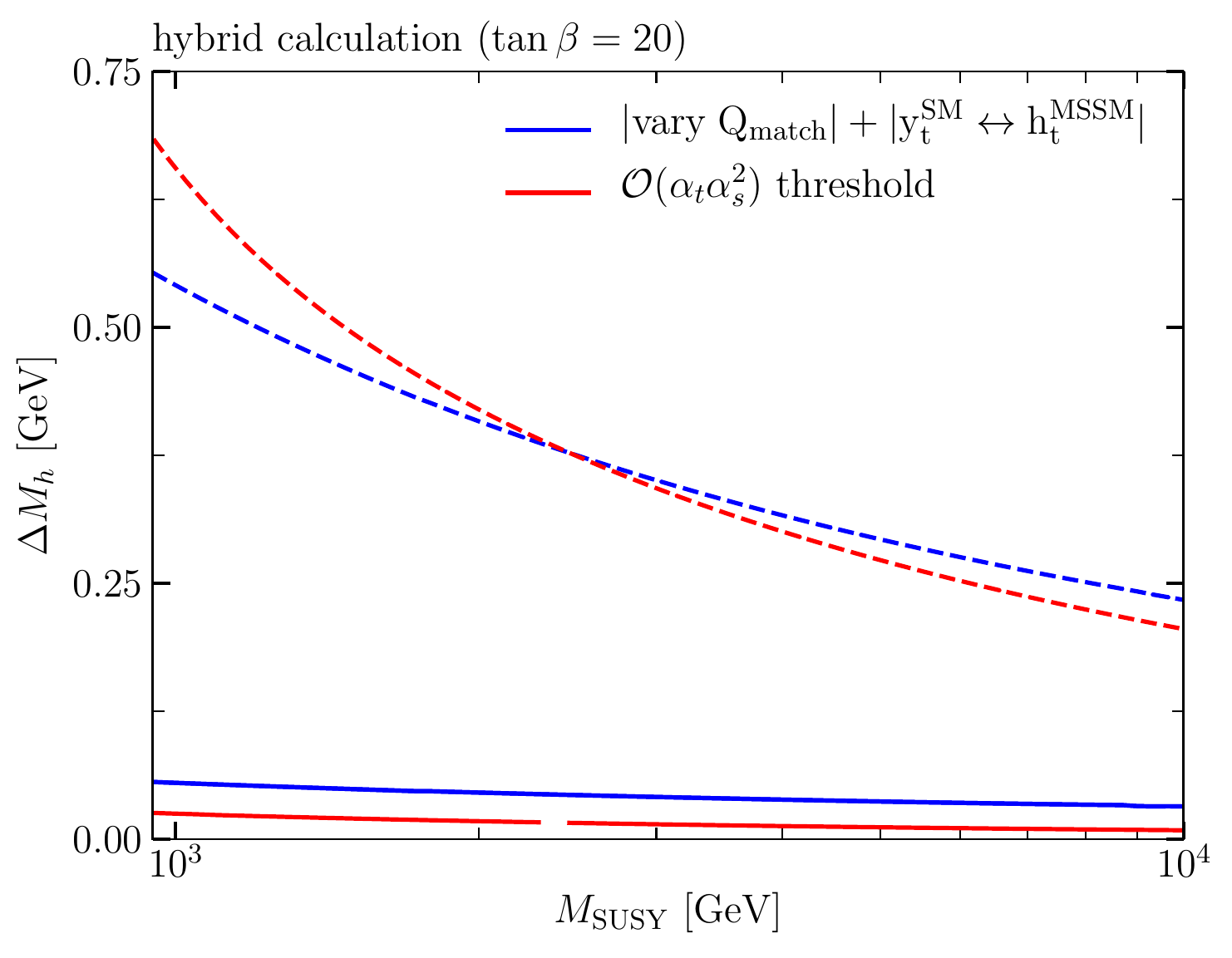}
\end{minipage}
\caption{Shift in $M_h$ induced by taking into account the \order{\alt\als^2} threshold correction (red) in comparison to the corresponding uncertainty estimate (blue). The results are shown as a function of $\msusy$ for ${X_t^\DR/\msusy = 0}$ (solid) and ${X_t^\DR/\msusy = -\sqrt{6}}$ (dashed), and for both cases ${\tan\beta = 20}$ is used. }
\label{fig:HSsrcHimalaya}
\end{figure}

In \Fig{fig:HSsrcHimalaya}, we cross-check our estimate of the high-scale uncertainty against the shift in $M_h$ induced by taking into account the \order{\alt\als^2} threshold correction provided by the publicly available code \texttt{Himalaya}~\cite{Harlander:2017kuc,Harlander:2018yhj}. For ${\xtDR=-\sqrt{6}}$, the induced shift from the \order{\alt\als^2} threshold correction varies between about 0.7 and 0.2~GeV for the displayed range of \msusy. The shift agrees well with our high-scale uncertainty estimate (within $\sim 0.2\gev$). For ${\xtDR=0}$ the shift induced by the threshold correction is much smaller, below 0.1~GeV. Also in this case it agrees very well with our high-scale uncertainty estimate. The small gap for ${\xtDR=0}$ and $\msusy\sim 2.5\tev$ is due to a numerical instability of \texttt{Himalaya} at this point. The comparison in \Fig{fig:HSsrcHimalaya} shows that our estimate of the high-scale uncertainty is indeed a good estimate for the possible numerical impact of higher-order threshold corrections. Also the shifts of $M_h$ induced by taking into account the mixed QCD-electroweak two-loop threshold correction~\cite{Bagnaschi:2019esc}, which is not included in our calculation, lie well within our uncertainty estimate.

\medskip

\begin{figure}\centering
\begin{minipage}{.48\textwidth}\centering
\includegraphics[width=\textwidth]{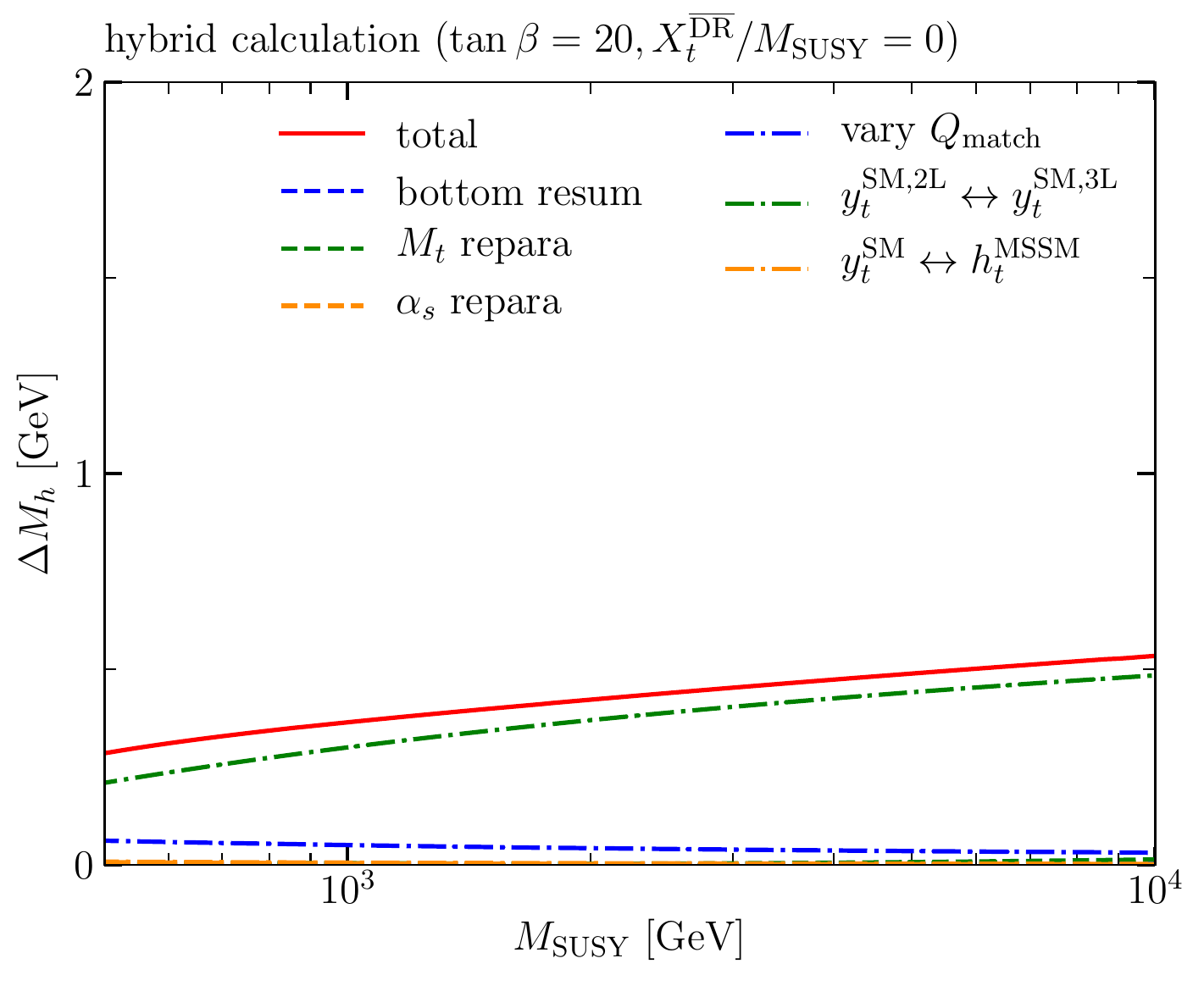}
\end{minipage}
\begin{minipage}{.48\textwidth}\centering
\includegraphics[width=\textwidth]{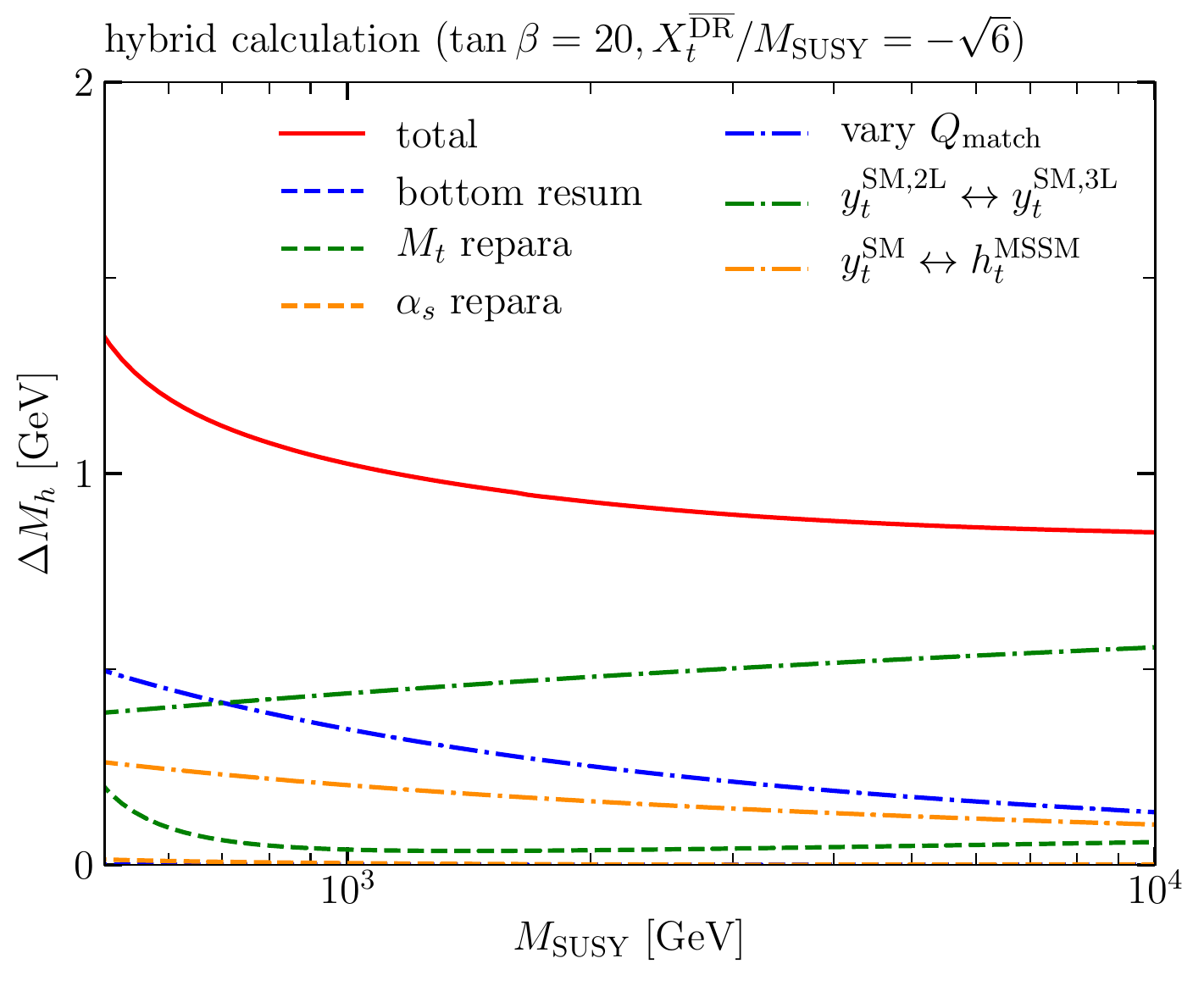}
\end{minipage}
\caption{Left: Uncertainty estimate of the hybrid calculation as a function of $\msusy$ for $\tan\beta = 20$ and $X_t^\DR/\msusy = 0$. Right: Same as left plot, but ${X_t^\DR/\msusy = -\sqrt{6}}$.}
\label{fig:HSsrcHybrid}
\end{figure}

Finally, we assess the uncertainty of the hybrid calculation restricting ourselves to the case of \DR renormalized stop parameters. In the left plot of \Fig{fig:HSsrcHybrid}, the corresponding uncertainty estimate is shown as a function of $\msusy$ for $X_t^\DR/\msusy = 0$. The total estimate (red) is almost constant (at $\sim 0.5\gev$) in the depicted $\msusy$ interval. We observe that the uncertainty estimates associated with the EFT calculation -- reparametrization of the threshold corrections in terms of the MSSM top Yukawa coupling (orange dot-dashed), variation of the matching scale (blue dot-dashed) and extracting the SM top Yukawa coupling at the three-loop instead of at the two-loop level (green dot-dashed) -- are by construction very similar to the case of the pure EFT calculation. The estimated uncertainty associated with terms that are suppressed by the SUSY scale was found to be relatively small for the no-mixing case in \Fig{fig:HSsrcEFT}. On the other hand, the uncertainty estimates having the biggest impact for the case of the pure fixed-order part -- replacing $\als$ by $\als\left[1 \pm \als/(4\pi)\ln(\msusy^2/M_t^2)\right]$ (orange dashed) and the reparametrization of the top-quark mass (gren dashed) -- yield contributions that are drastically reduced in comparison to the pure fixed-order part of the calculation with a numerical effect much below 0.1~GeV. This reflects the fact that the non-logarithmic terms are parametrized as in the EFT calculation and that consequently only terms suppressed by $\msusy$ are reparametrized. Here, it should be kept in mind that the size of unsuppressed higher-order terms proportional to the strong gauge coupling is assessed by the variation of the matching scale as well as the reparametrization of the top Yukawa coupling in case of SUSY contributions and the reparametrization of the top-quark mass in the pole determination in case of SM contributions.

The uncertainty estimate of the hybrid approach for $X_t^\DR/\msusy = -\sqrt{6}$ is shown in the right plot of \Fig{fig:HSsrcHybrid}. Because of the much bigger impact of the terms that are suppressed by the SUSY scale, see the right plot of \Fig{fig:HSsrcEFT}, in this case the uncertainty estimate for the result of the hybrid approach is reduced very significantly in comparison to both the result of the pure fixed-order part and the one of the pure EFT calculation. The reparametrization of the strong gauge coupling has again only a small numerical impact, for the same reason as explained above for the case of no mixing. The overall behaviour of the total uncertainty estimate is similar to the case of vanishing stop mixing resulting in a nearly constant estimate of $\sim 0.9\gev$ for $\msusy\gtrsim 1\tev$. The increase of the uncertainty estimate for low $\msusy$ is mainly caused by a larger uncertainty associated with higher-order threshold corrections. In addition, the uncertainty associated with terms suppressed by the SUSY scale beyond the order of the fixed-order calculation increases.

\medskip

\begin{figure}\centering
\begin{minipage}{.48\textwidth}\centering
\includegraphics[width=\textwidth]{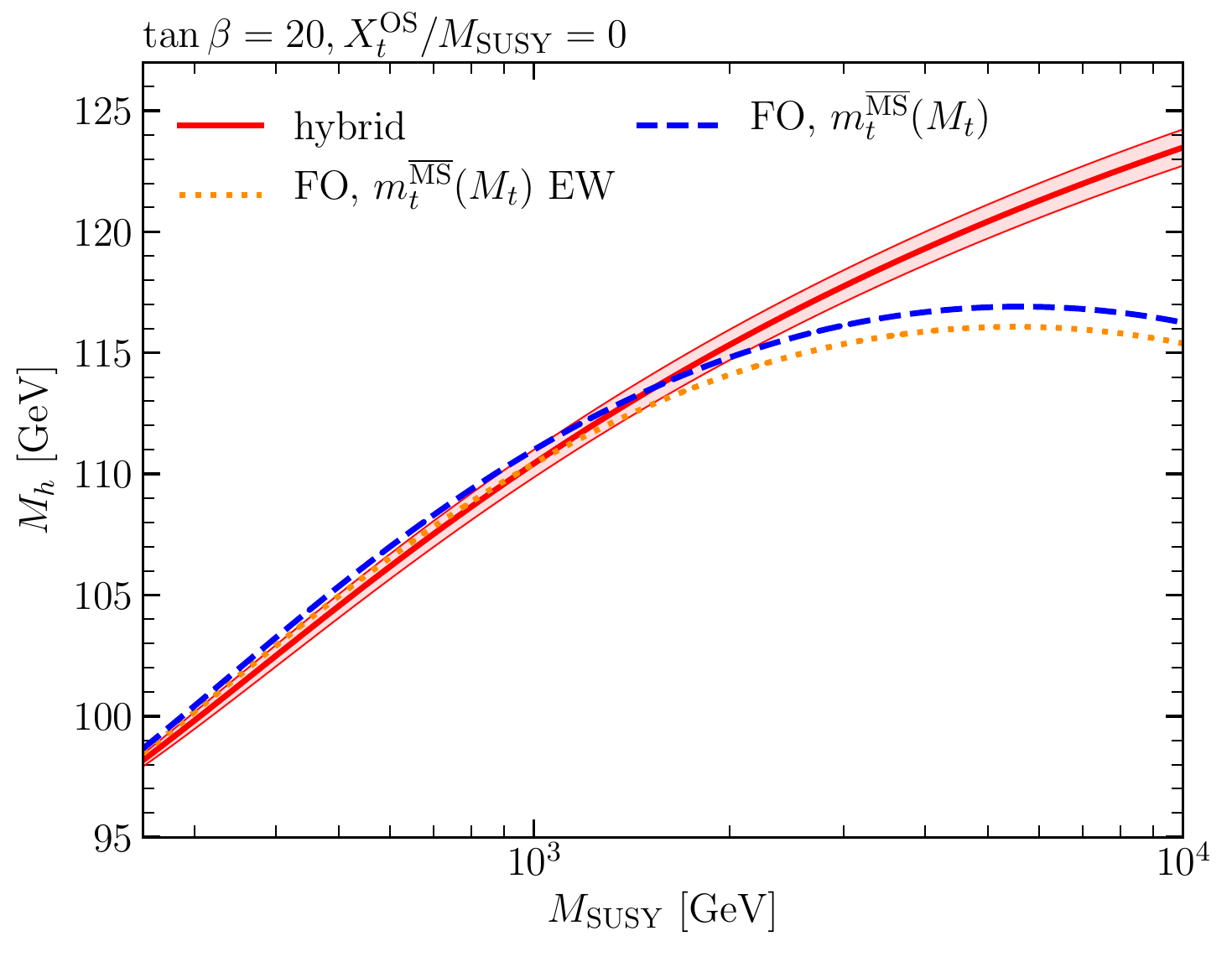}
\end{minipage}
\begin{minipage}{.48\textwidth}\centering
\includegraphics[width=\textwidth]{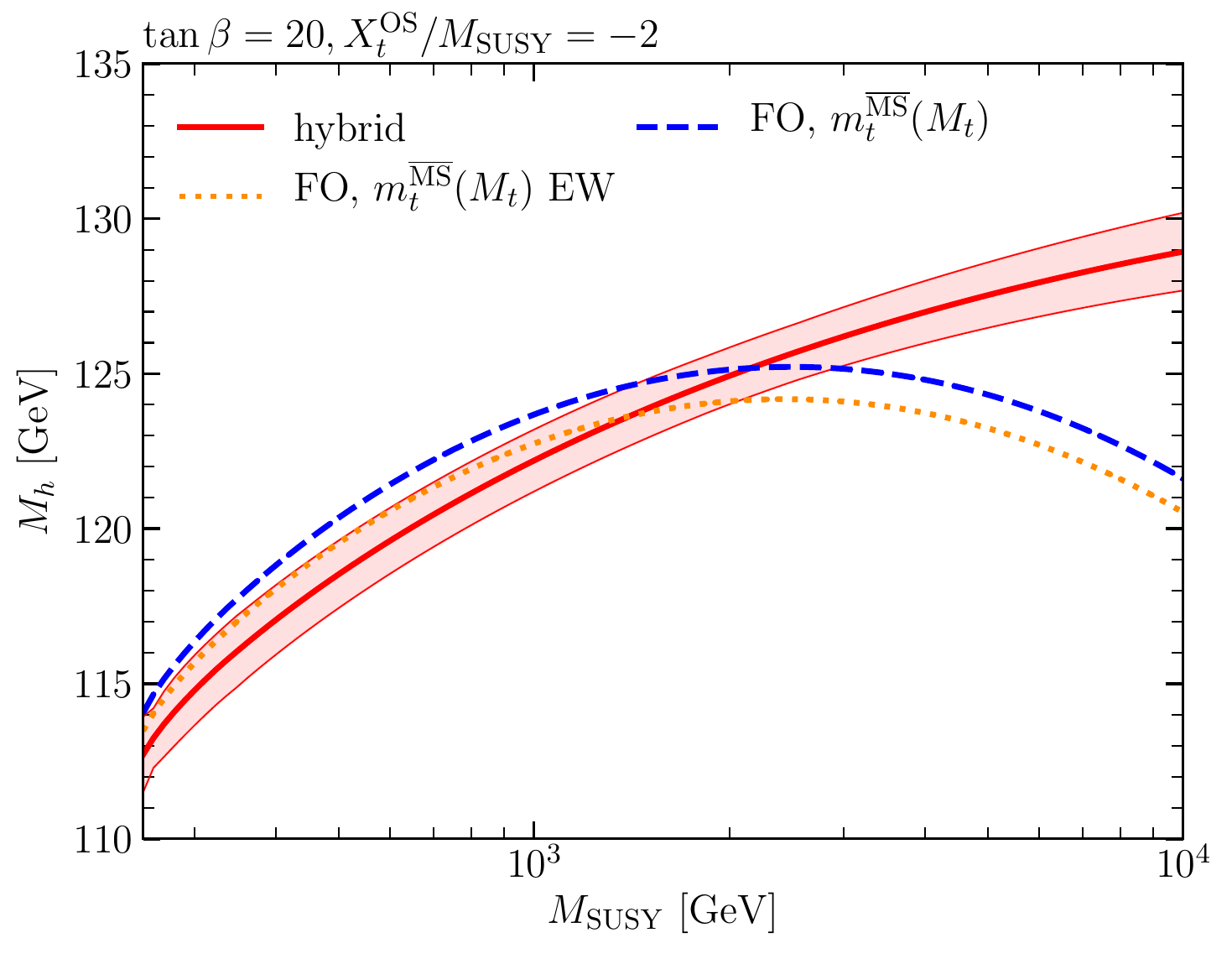}
\end{minipage}
\caption{Comparison of $M_h$ predictions as a function of $\msusy$ for $\tan\beta = 20$. The stop mixing parameter is chosen to be $X_t^\OS/\msusy = 0$ (left), and $X_t^\OS/\msusy = -2$ (right). The result obtained in the hybrid approach with associated uncertainty band (red) is compared with the fixed order result using the SM \MS top quark mass without (blue dashed) and with (orange dotted) electroweak one-loop corrections.}
\label{fig:HScomp_OS}
\end{figure}

After discussing the three different approaches and their associated uncertainty estimates individually up to now, we now compare the results of the three approaches with each other. We first focus on the case of OS input parameter in the stop sector in \Fig{fig:HScomp_OS}.\footnote{We do not show a pure EFT result here. Converting the OS input parameters first to the \DR scheme and then using these parameters as input for the EFT calculation would not be meaningful in this context as such a procedure would spoil the resummation of large logarithms (see the discussion in \cite{Bahl:2017aev}).} The predictions for $M_h$ are shown as a function of \msusy (thick lines). The colored red band depicts the uncertainty estimate for the hybrid result that has been obtained as described above. We do not display uncertainty estimates for the fixed-order results shown in \Fig{fig:HScomp_OS} (see the discussion of \Fig{fig:HSsrcFO_OS}). In addition to the default fixed-order result parametrized using the SM \MS top-quark mass evaluated at the scale $M_t$, where electroweak one-loop corrections are not included, we also show the fixed-order result using the SM \MS top mass including electroweak one-loop corrections (also evaluated at the scale $M_t$). The latter parametrization is employed in the hybrid result.

For the case of vanishing stop mixing (left plot), the predictions are in good agreement with each other for SUSY scales below $\sim 2\tev$. As discussed above, terms that are suppressed by \msusy have only a minor numerical impact for $\xtOS = 0$. The fixed-order calculation (blue dashed) agrees well with the logarithmic behaviour of the hybrid result (red solid) up to $\sim 2\tev$, while for larger values of \msusy an increasing discrepancy is visible. This deviation is caused by higher-order logarithmic contributions contained in the hybrid result that become large for increasing \msusy. Including electroweak corrections to the SM \MS top-quark mass in the fixed-order approach (orange dotted) yields sizeable shifts only above the TeV scale.

In case of $\xtOS = -2$, there is a larger deviation between the fixed-order and the hybrid result for SUSY scales below $\sim 2\tev$. The shift that persists down to very low SUSY scales, $\msusy\sim 250\gev$, can mainly be explained by the use of different \MS top-quark masses in the two results. The difference between the blue dashed and the orange dotted curves is caused by the electroweak one-loop corrections that shift the \MS top-quark mass downwards by $\sim 1\gev$ resulting in a downward shift of $M_h$ by also $\sim 1\gev$ (for more details see \cite{Bahl:2016brp}). The remaining difference between the hybrid result and the fixed-order result employing the SM \MS top mass including electroweak contributions is due to differences in the scheme choices of other parameters.\footnote{The fixed-order calculation is parametrized in terms of the SM \MS top mass and $v_{G_F}$, while in the hybrid calculation the SM \MS top Yukawa coupling and $v^\MS$ are employed (see also the discussion in \Sec{sec:02_HYBintro}). More details as well as numerical examples can be found in \cite{Bahl:2017aev}.}. The uncertainty associated with these different scheme choices is accounted for by the uncertainty estimate of the hybrid calculation, which is illustrated by the fact that the fixed-order result lies (just) within the uncertainty band of the hybrid calculation for the lowest displayed values of $\msusy$.

\begin{figure}\centering
\begin{minipage}{.48\textwidth}\centering
\includegraphics[width=\textwidth]{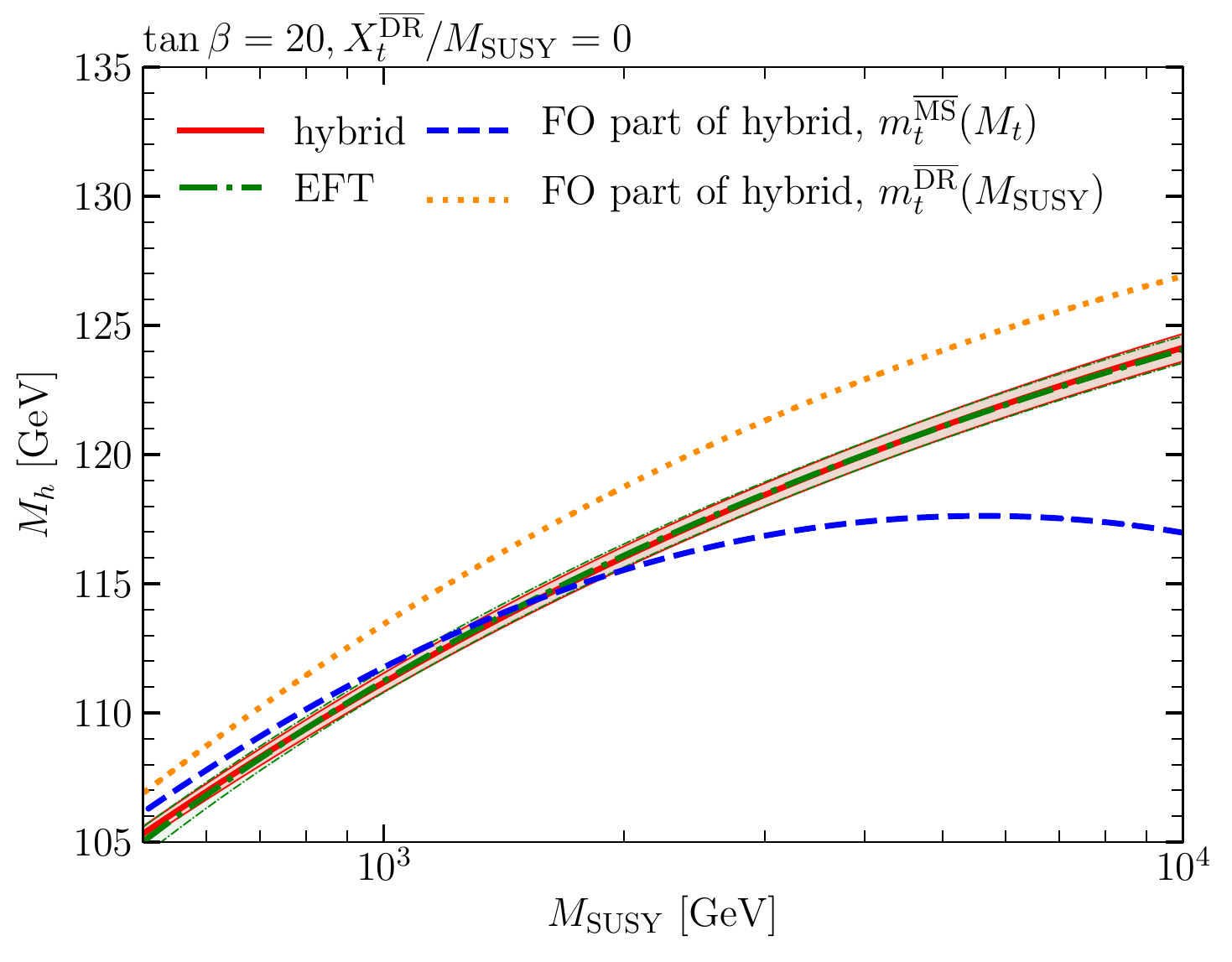}
\end{minipage}
\begin{minipage}{.48\textwidth}\centering
\includegraphics[width=\textwidth]{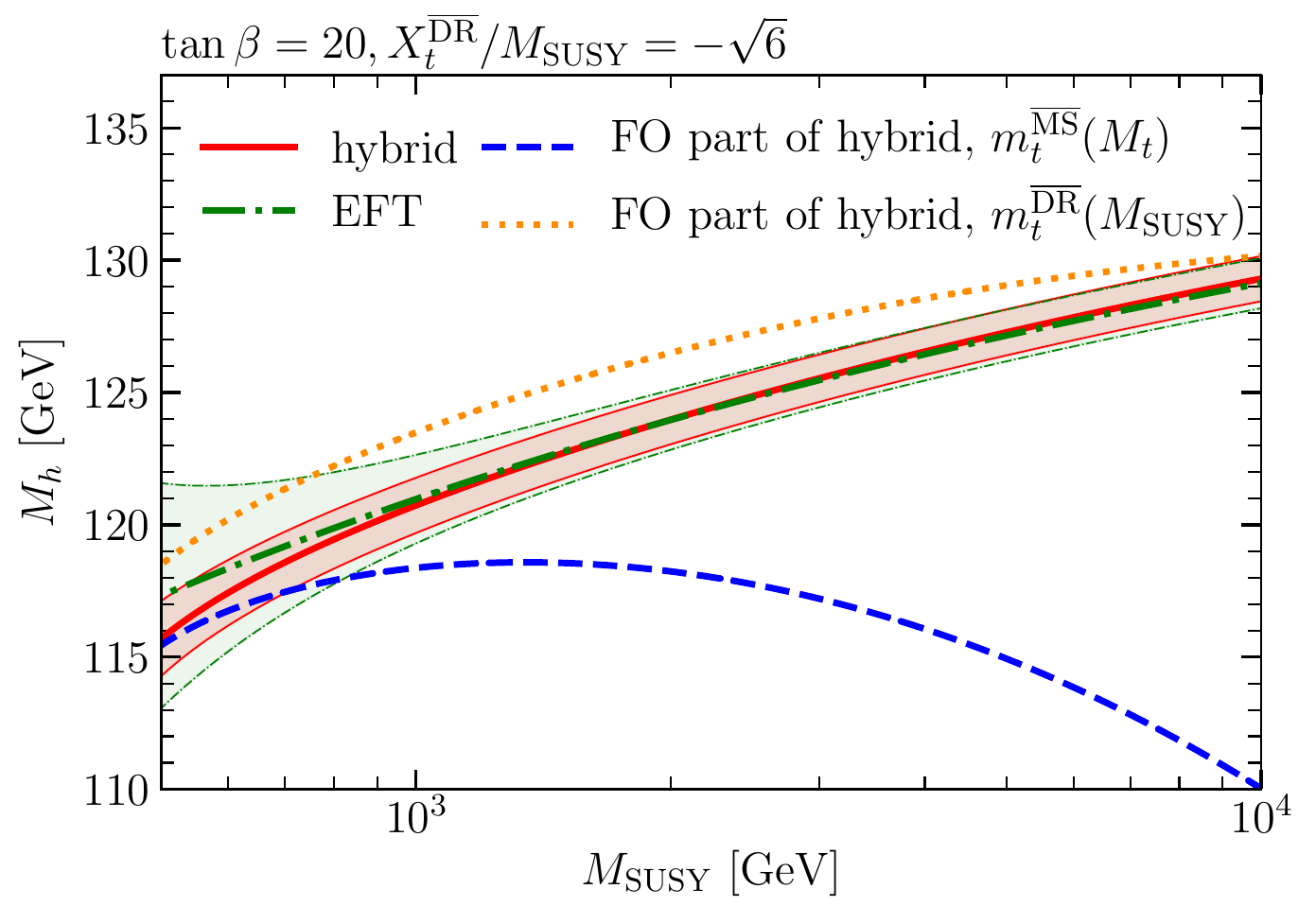}
\end{minipage}
\caption{Comparison of the $M_h$ predictions in the different approaches as a function of $\msusy$ for $\tan\beta = 20$. The stop mixing parameter is chosen to be $X_t^\DR/\msusy = 0$ (left) and ${X_t^\DR/\msusy = -\sqrt{6}}$ (right). The associated uncertainty band for the result of the hybrid approach (red) is compared with the one of the EFT result (green). The fixed-order part of the hybrid calculation is shown for the SM \MS top-quark mass evaluated at the scale $M_t$ (blue dashed) and for the MSSM \DR top-quark mass evaluated at the scale $\msusy$ (orange dotted).}
\label{fig:HScomp_DR}
\end{figure}

In \Fig{fig:HScomp_DR}, we compare the hybrid, the EFT and the fixed-order part of the hybrid calculation for the case of \DR input parameters in the stop sector. The case of $\xtDR=0$ (left plot) is similar to the case of vanishing stop mixing and OS input parameters in the stop sector. The uncertainty estimate of the hybrid calculation is, however, slightly reduced since no conversion of the stop input parameters between the OS and \DR scheme is necessary in this case (but as explained in \Sec{sec:05_HYBunc} additional theoretical uncertainties have to be taken into account in relating the \DR parameters to physical observables). In addition to the default fixed-order part of the hybrid calculation parameterized using the SM \MS top mass evaluated at the scale $M_t$, we also show the fixed-order part parameterized in terms of the MSSM \DR top mass evaluated at the scale $\msusy$ (orange dotted). We see that this result captures the logarithmic behaviour of the EFT and hybrid calculations up to much higher SUSY scales. It lies, however, above the EFT and the hybrid result by $\lesssim 3\gev$ in the whole considered $\msusy$ interval. Calculating the MSSM \DR top mass at a higher loop level (we only take into account one-loop corrections controlled by the strong gauge coupling and the top Yukawa coupling) would probably improve the agreement. The rather large difference between the two parameterizations of the fixed-order part for $\msusy \gtrsim 2\tev$ again indicates the importance of higher-order logarithmic contributions for large \msusy.

For the case of $\xtDR=-\sqrt{6}$ (right plot), the results of the EFT and the hybrid approach are in good agreement with each other except for low values of \msusy. As for the case $\xtOS = -2$ (see above), terms that are suppressed by \msusy have a much bigger numerical impact here than for the case of no mixing in the stop sector. This leads to a discrepancy between the EFT result, where these suppressed terms are not included, and the hybrid result for $\msusy\lesssim 2\tev$. The discrepancy amounts to about $1.5\gev$ in $M_h$ for $\msusy\sim 500\gev$. The uncertainty band of the hybrid approach is always smaller than the one of the EFT result for the whole displayed range of \msusy, which confirms that the hybrid approach yields a precise prediction for $M_h$ for all values of \msusy. As expected, the uncertainty band of the EFT result is much wider than the one of the hybrid result for $\msusy\lesssim 1\tev$ since terms that are suppressed by the SUSY scale are not included in the EFT result. The fixed-order part of the hybrid calculation using the SM \MS top mass agrees well with the hybrid result for small SUSY scales, $\msusy \sim 500\gev$, but it shows significant deviations from the hybrid result already for moderate SUSY values below $1\tev$ (which is related to the scale choices, see the discussion above). As for the case of no mixing, the logarithmic behaviour of the hybrid and EFT results can be captured up to much higher SUSY scales by parametrizing the fixed-order part in terms of the MSSM \DR top-quark mass at the scale $\msusy$, but this prediction is again significantly shifted upwards compared to the hybrid and EFT results (as explained above, the \DR top-quark mass is calculated including only one-loop corrections in the gaugeless limit).

\medskip

\begin{figure}\centering
\includegraphics[width=.7\textwidth]{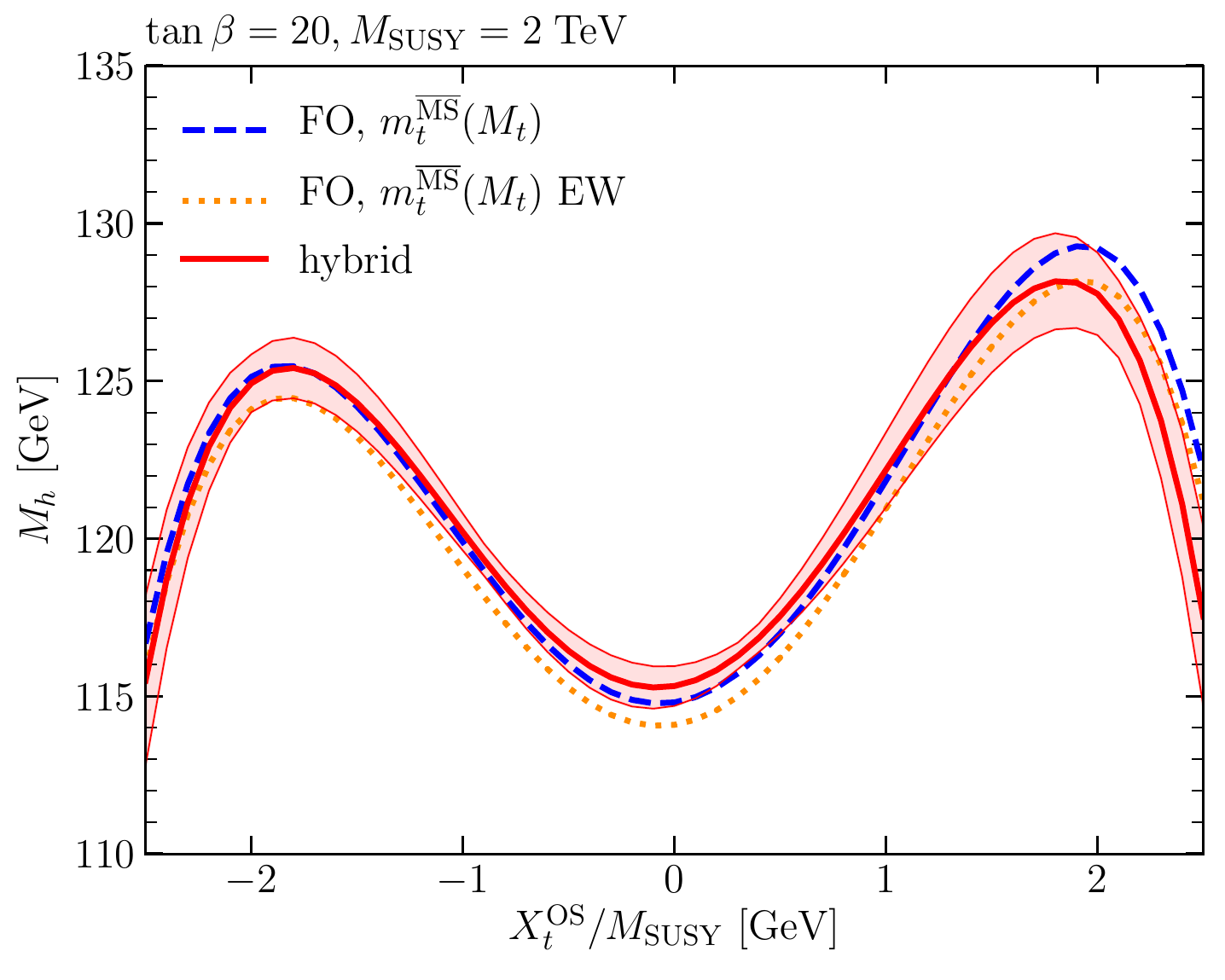}
\caption{Comparison of $M_h$ predictions as a function of $X_t^\OS/\msusy$ setting $\tan\beta = 20$ and ${\msusy = 2\tev}$. The result obtained in the hybrid approach with associated uncertainty band (red) is compared with the fixed-order result using the SM \MS top quark mass without (blue dashed) and with (orange dotted) electroweak one-loop corrections.}
\label{fig:varXtOS}
\end{figure}

In \Fig{fig:varXtOS}, we compare the hybrid (red) and the fixed-order approach, using the SM \MS top-quark mass without (blue dashed) and with (orange dotted) electroweak one-loop corrections, in the case of OS stop input parameters as a function of \xtOS fixing ${\msusy = 2\tev}$. The three results are seen to agree with each other with deviations of up to $\sim 1\gev$ for all displayed values of ${\xtOS\lesssim 2}$. In the interval $-2 \leq \xtOS \leq 2$, the uncertainty estimate for the hybrid approach is largest for ${\xtOS\simeq 1.6}$, where it amounts to $\Delta M_h = 2.3\gev$. For $\xtOS = 2$, we obtain ${\Delta M_h = 1.6\gev}$.

\begin{figure}\centering
\includegraphics[width=.7\textwidth]{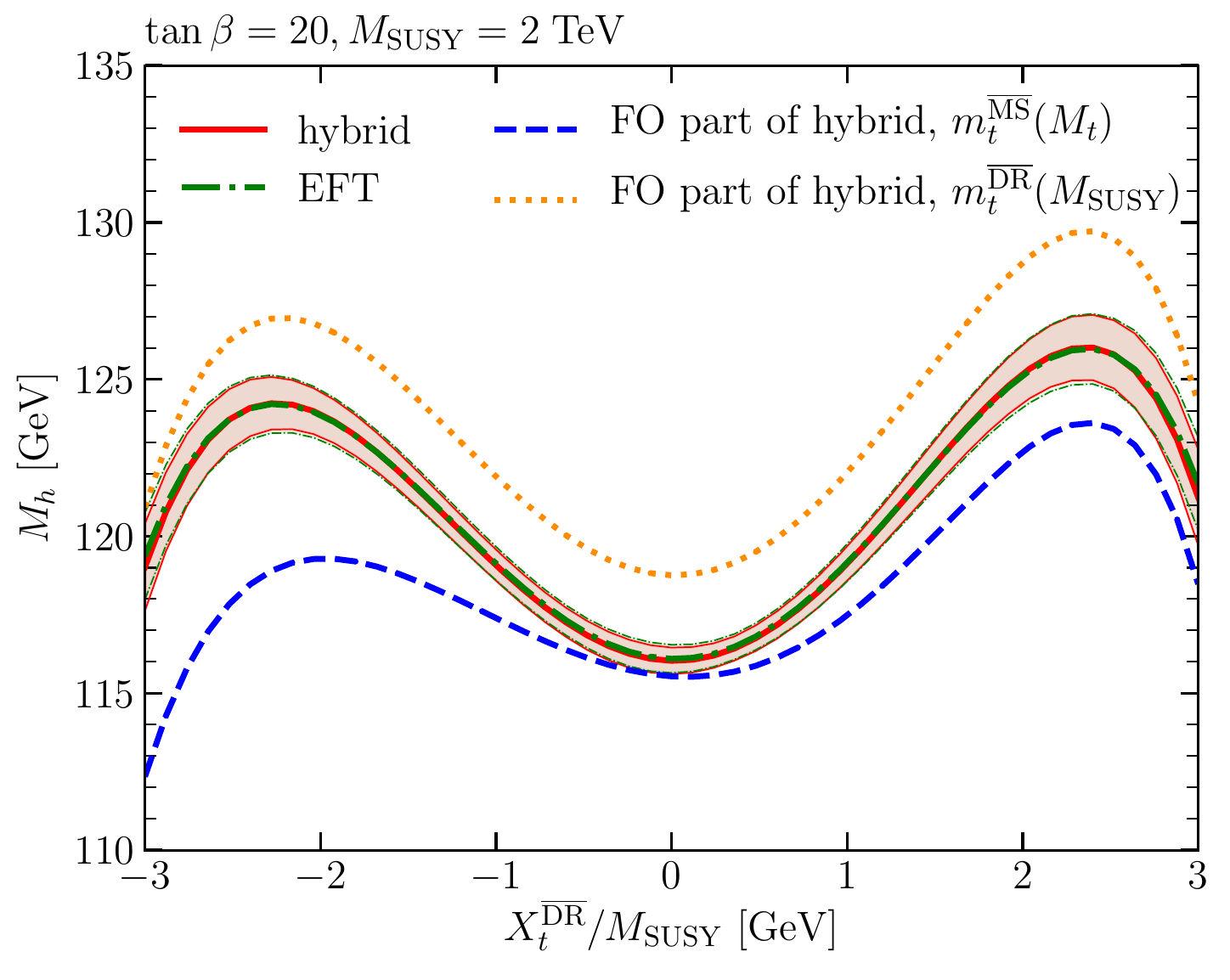}
\caption{Comparison of $M_h$ predictions as a function of $X_t^\DR/\msusy$ setting $\tan\beta = 20$ and ${\msusy = 2\tev}$. The result obtained in the hybrid approach with associated uncertainty band (red) is compared with the one of the EFT result (green). The fixed-order part of the hybrid calculation is shown for the SM \MS top-quark mass evaluated at the scale $M_t$ (blue dashed) and for the MSSM \DR top-quark mass evaluated at the scale $\msusy$ (orange dotted).}
\label{fig:varXtDR}
\end{figure}

In \Fig{fig:varXtDR}, we compare the hybrid, the EFT and the fixed-order part of the hybrid calculation with each other for the case of \DR input parameters, where $\xtDR$ is varied. The EFT (green dot-dashed) and the hybrid (red) results as well as their uncertainty estimates are in very good agreement with each other. In comparison to \Fig{fig:varXtOS}, the uncertainty band of the hybrid calculation is reduced by $\sim 1\gev$. This decrease arises since no conversion of $X_t$ from the OS scheme, as used in the fixed-order part of the calculation (in case of OS stop input parameters), to the \DR scheme, as used in the EFT part of the calculation (see the discussion in \Sec{sec:02_HYBintro}), is necessary in this case. As already noted in \Fig{fig:HScomp_DR}, the fixed-order part of the hybrid calculation parametrized in terms of the SM \MS top mass (blue dashed) deviates significantly from the hybrid (and EFT) result for negative $X_t^\DR$ (which is related to the scale choices, see above). The deviation is seen to be less pronounced for positive $X_t^\DR$. If the fixed-order part of the hybrid calculation is paramerized in terms of the MSSM \DR top-quark mass (orange dotted), its behaviour as a function of \xtDR resembles more closely the one of the hybrid and EFT results, while the sizable upward shift is related to the fact that the \DR top-quark mass in the displayed result is calculated including only one-loop corrections in the gaugeless limit, see the discussion of \Fig{fig:HScomp_DR}.

\medskip

From the fact that the uncertainty band of the hybrid result is smaller if its fixed-order part is expressed in terms of \DR parameters in the stop sector rather than OS parameters, as observed in Figs.~\ref{fig:HScomp_OS}--\ref{fig:varXtDR}, one should not conclude that the \DR renormalization scheme is superior to other schemes for obtaining a precise prediction for the mass of the SM-like Higgs boson. As discussed above, in order to obtain a meaningful prediction for the relation between the Higgs-boson mass and other physical observables it is inevitable to relate the involved Lagrangian parameters to physical observables. Since this part of the calculation is subject to additional theoretical uncertainties, one should assess the suitability of different renormalization schemes in the context of the overall uncertainty of the relation between physical observables. A detailed study of this issue is beyond the scope of the present paper. In this context it should also be noted that results expressed in terms of OS masses have often been found to be better behaved for the case of a sizable splitting between the masses than their counterparts that are expressed in terms of \DR masses (see also the discussion below and e.g.\ in \cite{Degrassi:2001yf}).

\medskip

\begin{figure}\centering
\begin{minipage}{.48\textwidth}\centering
\includegraphics[width=\textwidth]{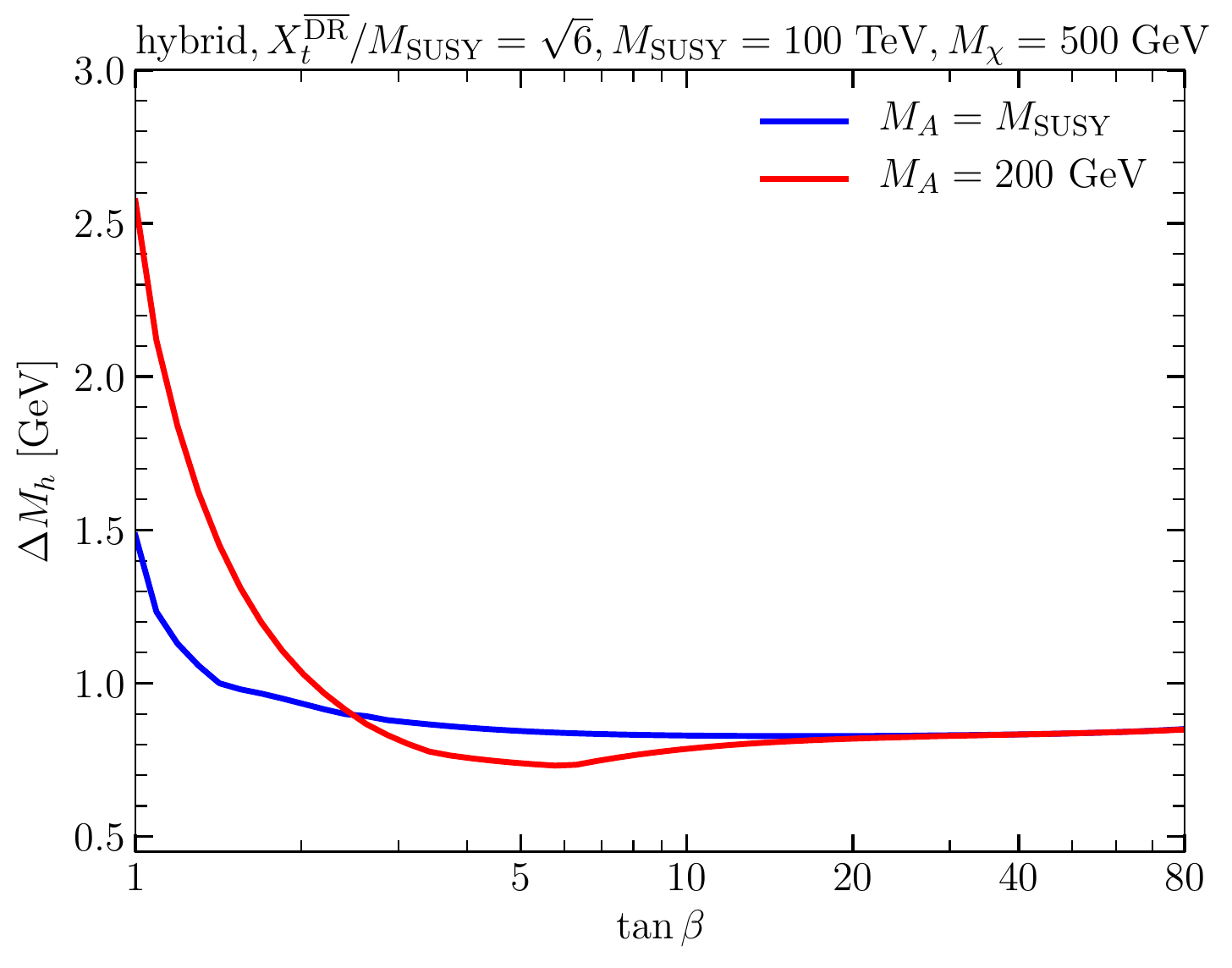}
\end{minipage}
\begin{minipage}{.48\textwidth}\centering
\includegraphics[width=\textwidth]{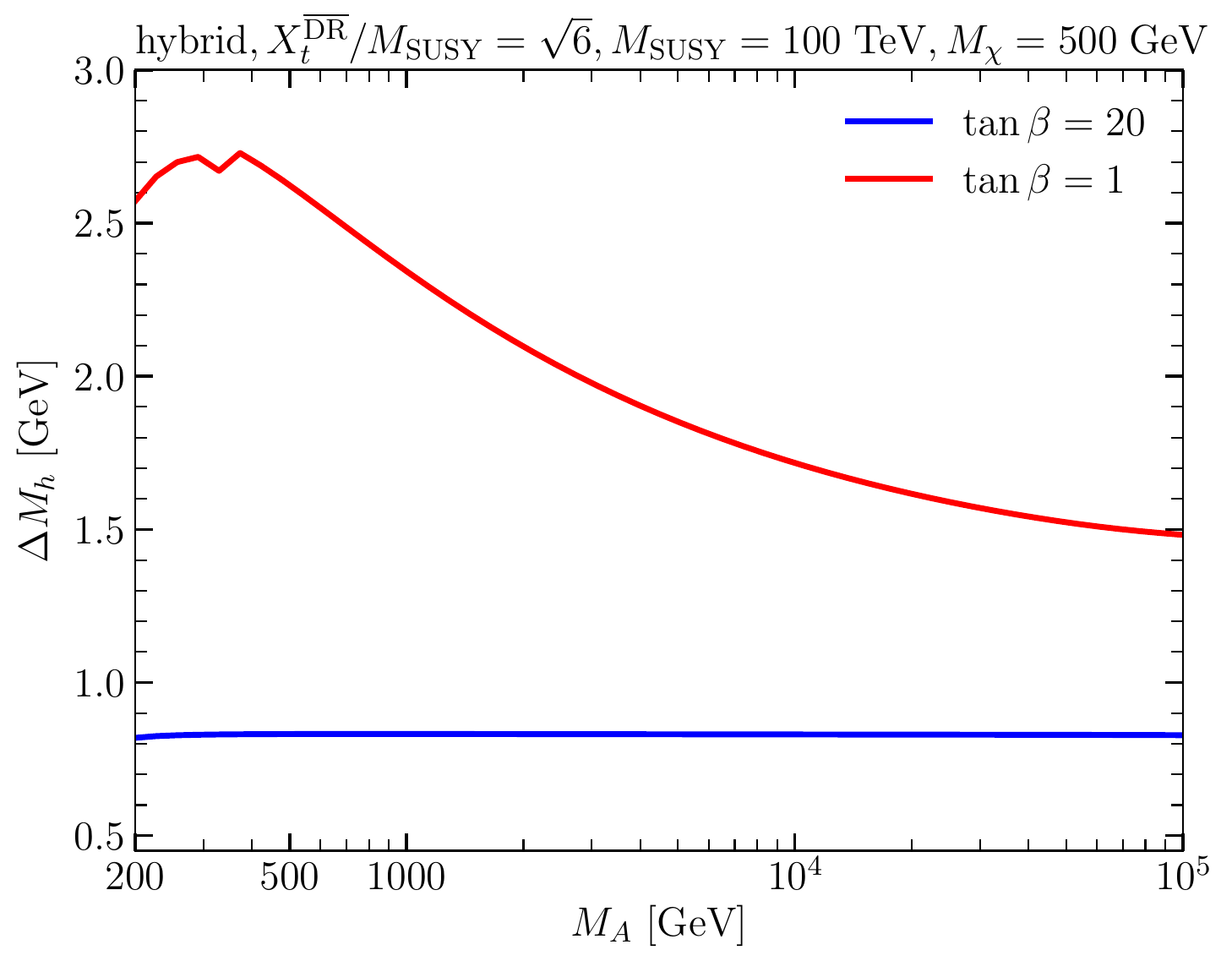}
\end{minipage}
\caption{Left: Uncertainty estimate of the hybrid calculation as a function of $\tan\beta$ setting $M_A = \msusy$ (blue) and $M_A = 200\gev$ (red) as well as $X_t^\DR/\msusy = \sqrt{6}$ and $\msusy = 100\tev$. Right: Same as left plot, but the uncertainty estimate is shown as a function of $M_A$ setting $\tan\beta = 20$ (blue) and $\tan\beta = 1$ (red).}
\label{fig:varTBMA}
\end{figure}

Next, we explore the dependence of the uncertainty estimate of our hybrid calculation on $\tan\beta$ and $M_A$. In order to introduce a large splitting between $M_A$ and $\msusy$ we use in this plot the very high value of $\msusy=100\tev$ (setting $\xtDR = \sqrt{6}$ and $\mu = M_1 = M_2 = 500 \gev$). This is the same scenario as the one that was used in Fig.~3 of~\cite{Bahl:2018jom} for comparing the Two-Higgs-Doublet-Model (THDM) and the SM as EFT below \msusy. The uncertainty estimate for the displayed result where the SM is used as effective low-energy theory can be compared with the size of the shift between the descriptions in terms of the THDM and the SM as low-energy EFT found in~\cite{Bahl:2018jom}. In the left plot of \Fig{fig:varTBMA}, we vary $\tan\beta$ setting $M_A = \msusy$ (blue) and $M_A = 200 \gev$ (red). In case of $M_A=\msusy$, the uncertainty estimate decreases from $\sim 1.5\gev$ for $\tan\beta = 1$ to $\sim 1\gev$ for $\tan\beta = 10$. This decrease reflects the fact that for low $\tan\beta$ the tree-level mass of the $h$ boson is smaller. Therefore, a fixed shift in the loop corrections becomes larger relative to the tree-level contribution. This leads to a larger shift in the final result for $M_h$ in case of low $\tan\beta$. For $\tan\beta \gtrsim 10$ the uncertainty estimate is approximately constant. For the case of $M_A = 200\gev$, the behaviour of the curve is similar to the case of $M_A = \msusy$. For $\tan\beta \sim 1$ the uncertainty estimate is, however, enlarged by $\sim 1\gev$. This increase originates mainly from Higgs mixing effects which are relevant in case of low $M_A$ and $\tan\beta$ (also the increased size of the MSSM top Yukawa coupling leads to an increase of the uncertainty).

These observations are corroborated by the results shown in the right plot of \Fig{fig:varTBMA} displaying the uncertainty estimate as a function of $M_A$ setting $\tan\beta = 20$ (blue) and $\tan\beta=1$ (red). We observe that the dependence on $M_A$ of the result for $\tan\beta = 20$ is negligible. For $\tan\beta = 1$, on the other hand, the uncertainty estimate decreases by $\sim 1.2\gev$ over the considered $M_A$ range. The small kink visible at $M_A\sim 340\gev$ is due to the mass of the $A$~boson passing the top-quark threshold ($M_A = 2 M_t$).

\Fig{fig:varTBMA} shows that the uncertainty estimate of the hybrid calculation with the SM as EFT below \msusy is found to be compatible with the differences found in Fig.~3 of~\cite{Bahl:2018jom} between the calculation using the SM as EFT and the calculation using the THDM as EFT, which should be more precise in the considered scenario due to the large mass hierarchy between $M_A$ and \msusy.

\medskip

\begin{figure}\centering
\begin{minipage}{.48\textwidth}\centering
\includegraphics[width=\textwidth]{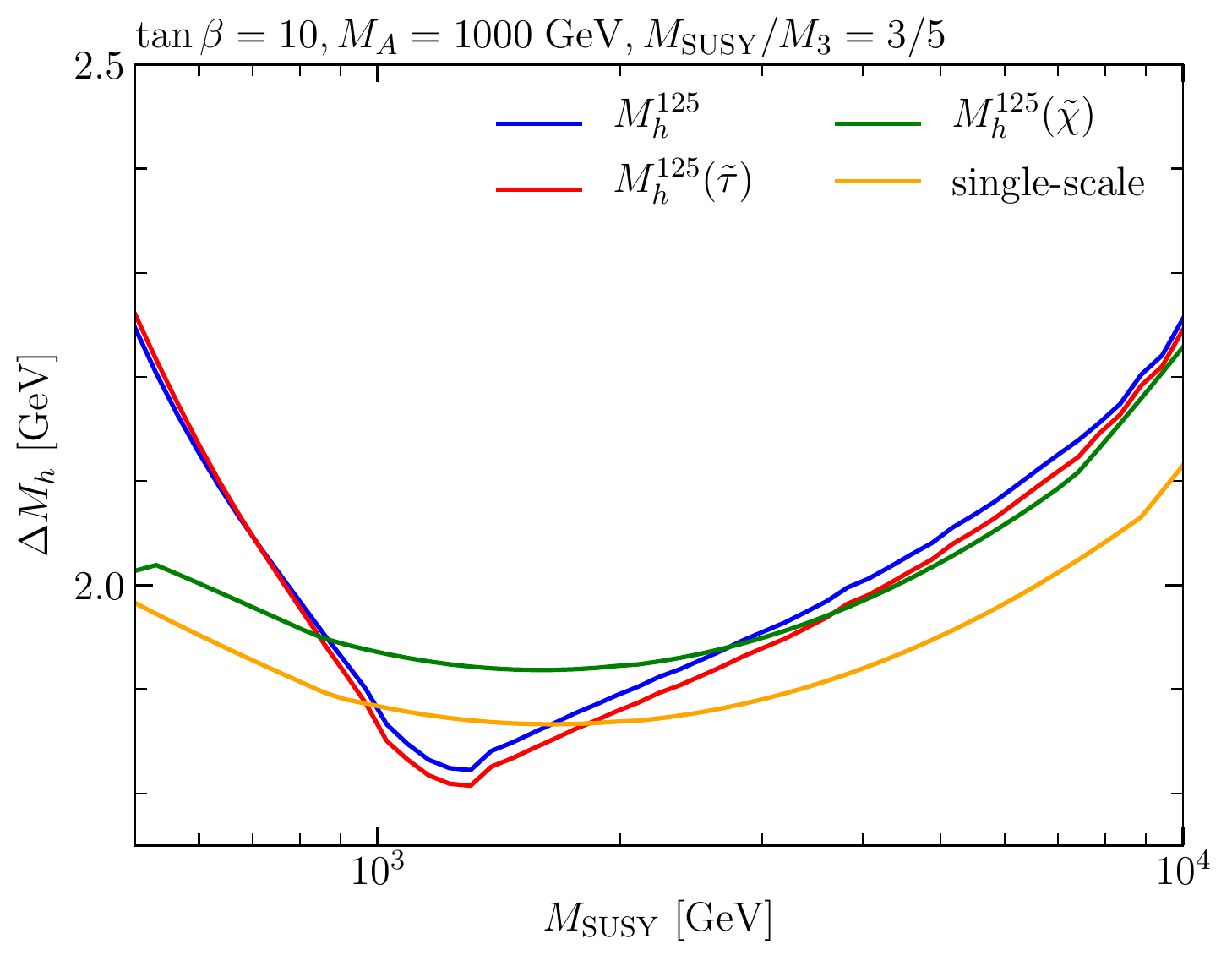}
\end{minipage}
\begin{minipage}{.48\textwidth}\centering
\includegraphics[width=\textwidth]{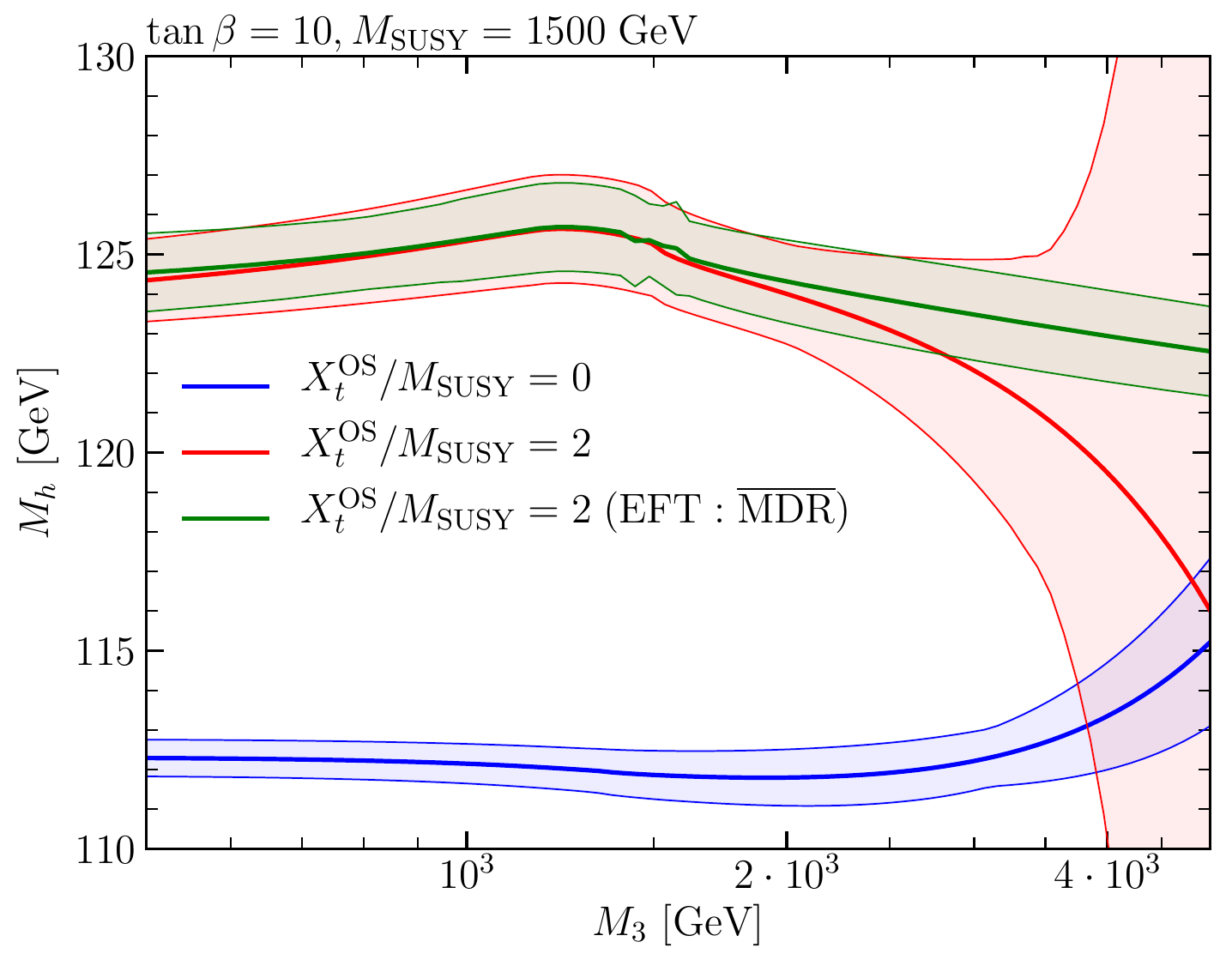}
\end{minipage}
\caption{Left: Uncertainty estimate of the hybrid calculation for some of the Higgs benchmark scenarios defined in~\cite{Bahl:2018zmf} as a function of $\msusy$ for $\tan\beta = 10$, $M_A = 1\tev$ and $\xtOS = 2$. For comparison also the result for a single-scale scenario is shown. For all four scenarios the gluino mass is fixed by imposing $\msusy/M_3 = 3/5$. Right: Results of the hybrid calculation for $M_h$ and the associated uncertainty estimate are shown as function of the gluino mass for $\xtOS = 0$ and $\xtOS = 2$. All other non-SM masses are fixed to $1.5\tev$. The green curve with its associated uncertainty band is generated employing the $\overline{\text{MDR}}$ scheme proposed in~\cite{Bahl:2019wzx} in the EFT part of the calculation.}
\label{fig:fitbenchmark}
\end{figure}

Up to now, we considered simplified scenarios with only one or two relevant mass scales. In order to investigate possible effects in scenarios with more complicated mass hierarchies, we explore in the left plot of \Fig{fig:fitbenchmark} the uncertainty estimates for three of the MSSM Higgs benchmark scenarios recently proposed in~\cite{Bahl:2018zmf}: the $M_h^{125}$ scenario with all SUSY particles above the TeV scale (blue), the $M_h^{125}(\tilde\tau)$ scenario featuring light staus as well as gauginos (red) and the $M_h^{125}(\tilde\chi)$ scenario featuring light gauginos and Higgsinos (green). In \cite{Bahl:2018zmf}, these scenarios were presented in $M_A$--$\tan\beta$ planes. As discussed above, the uncertainty estimate is only mildly dependent on $M_A$ and $\tan\beta$ for moderate $\tan\beta$ (the low $\tan\beta$-region is disfavoured for these scenarios by a too low $M_h$ prediction,\footnote{Estimating the theoretical uncertainty in the $M_{h,\rm{EFT}}^{125}$ scenario \cite{Bahl:2019ago}, in which the sfermion masses are pushed up to $10^{16}\gev$ in order to obtain $M_h\sim125\gev$ also for low $\tan\beta$, goes beyond the scope of the present paper.} while the high $\tan\beta$ region is significantly constrained by direct searches for heavy Higgs bosons). Therefore, we fix $M_A$ equal to 1~TeV as well as $\tan\beta=10$ and present the scenarios here as a function of the soft-SUSY breaking masses in the third generation scalar quark sector, which are chosen equal to each other in all three scenarios. We denote this mass scale by \msusy. In contrast to the results presented above, the first and second generation squarks, the gauginos, the Higgsinos as well as the sleptons have, however, masses that are not equal to \msusy. The mass splittings enter the threshold corrections in the form of logarithms of the masses over the matching scale. Larger mass splittings typically result in larger threshold corrections and therefore in a larger impact of the matching scale variation and the reparametrization of the top Yukawa coupling.

In each of the three scenarios, we set $\xtOS = 2$. This means that $M_h < 125\gev$  for $\msusy\lesssim 1.5\tev$, which is the value chosen in the benchmark scenarios, and $M_h > 125\gev$  for $\msusy\gtrsim 1.5\tev$. The gluino mass, $M_3$, is fixed by imposing $\msusy/M_3 = 3/5$ (such that for $\msusy=1.5\tev$ the gluino mass chosen for the actual benchmark scenarios, $M_3 = 2.5\tev$, is reached). For comparison we also show the result for a single-scale scenario (orange) where all sfermions as well as the electroweak gauginos and Higgsinos share a common mass scale (we, however, still impose $\msusy/M_3 = 3/5$).

The curves in the left plot of \Fig{fig:fitbenchmark} displaying the uncertainty estimates in the four scenarios agree quite well with each other. The uncertainty estimate for the single-scale scenario is lower than for the other scenarios, except for the $M_h^{125}$ and $M_h^{125}(\tilde\tau)$ scenarios at $\msusy\sim 1\tev$ where an accidental cancellation within one of the ingredients of the uncertainty estimates results in a lower uncertainty estimate. On the other hand, for $\msusy\lesssim 1\tev$, the uncertainty estimates of the $M_h^{125}$ and $M_h^{125}(\tilde\tau)$ scenarios are slightly larger (up to $\sim 0.3\gev$) in comparison to the other two scenarios, since in the $M_h^{125}$ and $M_h^{125}(\tilde\tau)$ scenarios $\mu/\msusy$ becomes larger than unity resulting in an increased size of terms that are formally suppressed by $\msusy$. For $\msusy\gtrsim 1\tev$, the uncertainty estimates of all four scenarios increase in a similar way from $\sim 1.9\gev$ to $\sim 2.2\gev$ for $\msusy = 10\tev$. This increase, being in contrast to the behaviour found for raising $\msusy$ above, is caused by the hierarchy between the gluino mass and the other SUSY particles.

We investigate this behaviour further in the right plot of \Fig{fig:fitbenchmark}. In this plot, we set all non-SM masses equal to $1.5\tev$ except for the gluino mass, which we vary between $500\gev$ and $5\tev$. For $\xtOS=0$ (blue), the dependence of the $M_h$ prediction (thick line) and its uncertainty estimate (coloured band) on $M_3$ is relatively small. Only for $M_3\gtrsim 4\tev$ a moderate increase in $M_h$ and also of the corresponding uncertainty estimate is visible. For $\xtOS=2$ (red), however, there is a strong dependence of the prediction for $M_h$ on $M_3$ as soon as $M_3 > \msusy$ originating from terms growing approximately quadratic with the gluino mass (see \cite{Degrassi:2001yf}). In the hybrid result these are terms entering at the three-loop level via the EFT calculation, while as shown in~\cite{Heinemeyer:1998np} the two-loop fixed-order result with on-shell renormalization in the stop sector does not give rise to a quadratic (and also not a linear) dependence on $M_3$. For $M_3 \gtrsim 2\msusy$, also the uncertainty estimate shows a strong dependence on $M_3$ rising quickly from $\sim 1.5\gev$ (for $M_3\lesssim 2.2\tev$) to values above $10~\gev$ (for $M_3\sim 4\tev$). This sharp increase originates from the uncertainty associated with higher-order threshold corrections. For the variation of the matching scale the SUSY soft-breaking parameters, normally entering at the scale $\msusy$, have to be evolved to the scale $2\msusy$ (or $1/2\msusy$). The gluino mass parameter enters the RGE of the stop mixing parameter, $X_t$. In case of a large $M_3$, this leads to large $|X_t/\msusy|$ values. Since the $M_h$ prediction depends very strongly on $X_t/\msusy$ for values beyond the maxima, the RGE evolution of $X_t$ leads to large shifts in the $M_h$ prediction resulting in a correspondingly large uncertainty estimate. We also checked that using $X_t$ as input parameter in the \DR scheme instead of the OS scheme does not alleviate this issue.

The actual reason for the observed behaviour of the uncertainty estimate is the missing EFT prescription for the case where the gluino is heavier than the remaining SUSY particles. Whereas the EFT calculation, implemented in \FH, resums large logarithms associated with the gluino for $M_3 < \msusy$ (SM plus gluino as EFT, see \cite{Bahl:2016brp,Bahl:2018qog} for more details), the corresponding logarithms are not resummed for $M_3 > \msusy$ (and there are also non-logarithmic terms giving rise to the approximately quadratic dependence on $M_3$). The suitable EFT -- MSSM without gluino -- has not been worked out yet for the calculation of the SM-like Higgs-boson mass but it has been employed for the calculation of other observables \cite{Muhlleitner:2008yw,Aebischer:2017aqa,Kramer:2019fwz}, in which similar non-decoupling effects arise.

In order to stabilize the Higgs-mass prediction in case of a larger hierarchy between the stop masses and the gluino mass, using the MSSM without gluino as EFT is not mandatory. Instead, the $M_3$-enhanced terms can be absorbed into the stop soft SUSY-breaking parameters. This scheme, called $\overline{\text{MDR}}$ scheme, can be used in the EFT calculation to completely circumvent the appearance of $M_3$-enhanced terms in the hybrid calculation. A detailed discussed can be found in in~\cite{Bahl:2019wzx}. The green curve with its associated uncertainty band in the right plot of \Fig{fig:fitbenchmark} shows the result of this method for $X_t^\OS/\msusy = 2$. The Higgs-mass prediction is clearly stabilized for large $M_3/\msusy$, and the uncertainty estimate is drastically reduced in comparison to the result using the \DR scheme for the definition of the stop parameters in the EFT part of the calculation. The implementation based on the $\overline{\text{MDR}}$ scheme will become publicly available in an upcoming \FH version.

\medskip

For the benchmark scenarios investigated in \Fig{fig:fitbenchmark} we have found that the uncertainty estimate in the hybrid approach only mildly depends on the presence of additional mass scales, reflecting the fact that the prediction for the SM-like Higgs-boson mass typically has a pronounced dependence only on the parameters of the stop sector as well as on $\tan\beta$ and $M_A$. An important exception that we have pointed out is the case of a gluino that is significantly heavier than the other SUSY particles, where large unresummed logarithms as well as terms growing approximately linear with the gluino mass can lead to a very large uncertainty estimate. In general the uncertainty estimate for the prediction of the mass of the SM-like Higgs boson that is based on a given spectrum of the particle masses used as input for the calculation needs to take into account the uncertainty that is related to assigning a suitable EFT description to the given pattern of the mass spectrum. If hierarchies of scales are present for which no appropriate EFT description is available, the theoretical uncertainties can significantly increase, as we have demonstrated for the case of a heavy gluino. A detailed analysis of possible patterns of hierarchies that can occur in phenomenologically relevant models is beyond the scope of the present paper.


\section{Conclusions}
\label{sec:07_conclusions}

In this paper we have performed an estimate of the remaining theoretical uncertainties in the prediction for the SM-like Higgs-boson mass in the MSSM as implemented in \FH. The prediction is based on a hybrid result combining contributions that are obtained via the fixed-order and via the EFT approach. In order to assess the theoretical uncertainties from unknown higher-order corrections of the hybrid result, we have also discussed the uncertainties of the fixed-order and the EFT approach separately. The additional theoretical uncertainties that are induced by the experimental errors of the input parameters, for instance the experimental values of the top-quark mass and $\alpha_s(M_Z)$, can directly be assessed by varying those input parameters and have not been included in our discussion. We have emphasized, however, that in order to obtain a meaningful prediction for the relation between the Higgs-boson mass and other physical observables the set of all input parameters of the Higgs-mass prediction needs to be related to a corresponding set of physical observables (this applies in particular to the case where \DR parameters are taken as input parameters for the Higgs-boson mass prediction). As a consequence, additional theoretical uncertainties occur in the relation between the predicted Higgs-boson mass and other physical observables from relating the involved \DR parameters to suitable observables (the same applies to all other input parameters that are not expressed in terms of physical observables).

Our assessment of the theoretical uncertainties of the hybrid result applies to the implementation in \FH, where either all SUSY particles are integrated out at a common scale and the EFT at low scales is the SM, or the low-energy EFT is the SM with added gauginos and Higgsinos.\footnote{We leave an assessment of the uncertainties in case of the THDM as low-energy EFT (as implemented e.g.\ in \FH (version \texttt{2.16.0}) based on~\cite{Bahl:2018jom}) for future work.} In the employed EFT no logarithms proportional to the bottom Yukawa coupling are resummed. The uncertainty estimates obtained in this paper apply to the case of real MSSM parameters and invoke minimal flavour violation. For most of the numerical analysis a simple single-scale scenario has been chosen, in which all non-SM masses are taken to be equal to $\msusy$. The case of different SUSY scales has been investigated for a set of recently proposed benchmark scenarios for the search for additional Higgs bosons.

We have estimated the uncertainty of the pure fixed-order approach by employing different renormalization schemes for the top-quark mass, by multiplying the two-loop \order{\alt\als} correction with $\pm\als/(4\pi)\ln(\msusy^2/M_t^2)$ and by switching on and off the resummation of the bottom Yukawa coupling. If the OS scheme is employed for the renormalization of the stop sector, our estimate yields an uncertainty of the fixed-order approach of $\gtrsim 2\gev$ for SUSY masses above the TeV scale. For the case where the fixed-order contribution is expressed in terms of \DR stop input parameters, which appears in \FH only as part of the combined hybrid calculation, the estimate yields somewhat larger uncertainties already for relatively low values of $\msusy$ if there is large stop mixing. This behaviour can be traced to the different scale choices for the top-quark mass and the stop parameters in this part of the result (it could be remedied by using the MSSM \DR top-quark mass evaluated at the scale \msusy in this part of the result instead of the SM \MS top-quark mass evaluated at the scale $M_t$, which is used by default).

We have estimated the uncertainty of the pure EFT approach by assessing the uncertainty entering at the matching scale, the uncertainty entering at the electroweak scale and the uncertainty associated with terms suppressed by the SUSY scale separately. The resulting estimate is relatively low ($\sim 1\gev$) for SUSY masses above the TeV scale, since all relevant large logarithms are resummed for the considered single-scale scenario. For low values of the SUSY scale, however, the uncertainty of the EFT approach is increased, reaching values above $4\gev$ for large stop mixing, since terms which would be suppressed in case of a large SUSY scale can become relevant if no higher-dimensional operators are included. As an alternative estimate of the high-scale uncertainty of the EFT calculation, we compared our estimate to the shift induced by taking into account \order{\alt\als^2} threshold corrections computed by the public code \texttt{Himalaya} finding good agreement for large stop mixing.

Concerning the uncertainty of the hybrid calculation, we showed that the combination of the fixed-order and the EFT calculation leads to a reduction of different sources of uncertainty of the individual calculations. In the limit of low SUSY scales, the uncertainty estimate of the hybrid calculation approaches the one of the fixed-order calculation; in the limit of high SUSY scales, it approaches the one of the EFT calculation. For intermediary scales, we found the uncertainty estimate of the hybrid calculation to be below the estimates for the individual fixed-order and EFT calculations.

For the single-scale scenario and \DR input parameters in the stop sector our estimate yields a theoretical uncertainty from unknown higher-order corrections in the prediction for the SM-like Higgs boson mass of $\sim 0.5\gev$ for small stop mixing and up to $\sim 1.5\gev$ for large stop mixing. For on-shell input parameters in the stop sector the additional parameter conversion increases the uncertainty estimate by up to $\sim 1\gev$ for large stop mixing. These numbers can serve as a ``rule of thumb`` for the remaining theoretical uncertainties from unknown higher-order corrections.

As stressed above, the prediction expressed in terms of \DR input parameters is subject to additional theoretical uncertainties arising from the relation of the input parameters to physical observables. In this context we have also studied the dependence of the uncertainty estimate of the hybrid calculation on other relevant input parameters. We found that the uncertainty is somewhat enlarged for low $\tan\beta$ and low $M_A$.

Going beyond the case of a simple single-scale scenario, we also investigated some of the recently proposed Higgs benchmark scenarios for the search for additional Higgs bosons where some splitting occurs in the SUSY mass spectra. We found that the presence of those mass hierarchies did not significantly alter the behaviour of the uncertainty estimate in the hybrid approach with the exception of the case where the gluino mass, $M_3$, is larger than the remaining SUSY masses. This mass hierarchy is currently not covered by the EFT calculation leading to large uncertainties in the region of $M_3 \gtrsim 2\msusy$. We leave an improvement of the EFT calculation for this mass hierarchy for future work.

The presented uncertainty estimate is part of the publicly available code \FH from version \texttt{2.15.0} onwards.




\section*{Acknowledgments}
\sloppy{
We thank P.~Slavich for useful discussions. The work of S.H.\ is supported in part by the MEINCOP Spain under contract FPA2016-78022-P, in part by the ``Spanish Agencia Estatal de Investigaci\'on'' (AEI) and the EU ``Fondo Europeo de Desarrollo Regional'' (FEDER) through the project FPA2016-78022-P, in part by the ``Spanish Red Consolider MultiDark'' FPA2017-90566-REDC, and in part by the AEI through the grant IFT Centro de Excelencia Severo Ochoa SEV-2016-0597. H.B.\ and G.W.\ acknowledge support by the Deutsche Forschungsgemeinschaft (DFG, German Research Foundation) under Germany's Excellence Strategy -- EXC 2121 ``Quantum Universe'' -- 390833306.
}


\appendix



\newpage

\bibliographystyle{JHEP}
\bibliography{bibliography}{}

\end{document}